\documentclass[12pt,a4paper,final]{iopart}
\usepackage{graphicx}

\usepackage{lineno} 
\modulolinenumbers[5]
\usepackage{bm}
\usepackage[binary-units=true]{siunitx}

\newcommand{\voroxx}{{Voro++}{\,}{\,}}
\newcommand{\matlab}{{MATLAB}{\,}{\,}}
\newcommand{\dm}{{DAMASK}{\,}{\,}}

\newcommand{\ervat}[1]{$\overline{\epsilon_{vM}} = $ #1}
\newcommand{\mNull}{${\mathcal{P}}_n^0\,$\,}
\newcommand{\mNulleps}{${\mathcal{P}}_n^{0+\epsilon}\,$\,}
\newcommand{\mOne}{${\mathcal{P}}_n^1\,$\,}
\newcommand{\kNull}{${\mathcal{P}}_{k,n}^0\,$\,}

\newcommand{\kOne}{${\mathcal{P}}_{k,n}^1\,$\,}

\usepackage{booktabs}
\usepackage{tabularx}
\usepackage{subfig}
\usepackage{color}
\usepackage{iopams}

\usepackage{algorithm}
\usepackage[noend]{algpseudocode}
\def\BState{\State\hskip-\ALG@thistlm}
\algdef{SE}[DOWHILE]{Do}{doWhile}{\algorithmicdo}[1]{\algorithmicwhile\ #1}

\usepackage[cmyk]{xcolor}
\definecolor{mkre}{cmyk}{0.00,0.92,1.00,0.00}
\definecolor{mkgr}{cmyk}{0.60,0.00,1.00,0.00}
\definecolor{mklb}{cmyk}{0.51,0.47,0.00,0.00}
\definecolor{mkor}{cmyk}{0.00,0.20,0.85,0.00}
\definecolor{mkdb}{cmyk}{1.00,0.98,0.21,0.33}

\newcommand{\debug}[1]{\textcolor{mkre}{\textbf{ #1 }}}

\usepackage{cite}

\begin{document}

\title[]{Quantification of 3D spatial correlations between state variables and distances to the grain boundary network in full-field crystal plasticity spectral method simulations}
\author{Markus K\"{u}hbach, Franz Roters}

\address{Max-Planck-Institut f\"ur Eisenforschung GmbH, \\ Max-Planck-Str. 1, D-40237 D\"usseldorf}
\eads{\mailto{m.kuehbach@mpie.de}}

\begin{abstract}

Deformation microstructure heterogeneities play a pivotal role during dislocation patterning and interface network restructuring. Thus, they affect indirectly how an alloy recrystallizes if at all. Given this relevance, it has become common practice to study the evolution of deformation microstructure heterogeneities with 3D experiments and full-field crystal plasticity computer simulations including tools such as the spectral method.

Quantifying material point to grain or phase boundary distances, though, is a practical challenge with spectral method crystal plasticity models because these discretize the material volume rather than mesh explicitly the grain and phase boundary interface network. This limitation calls for the development of interface reconstruction algorithms which enable us to develop specific data post-processing protocols to quantify spatial correlations between state variable values at each material point and the points' corresponding distance to the closest grain or phase boundary.

This work contributes to advance such post-processing routines. Specifically, two grain reconstruction and three distancing methods are developed to solve above challenge. The individual strengths and limitations of these methods surplus the efficiency of their parallel implementation is assessed with an exemplary \dm large scale crystal plasticity study. We apply the new tool to assess the evolution of subtle stress and disorientation gradients towards grain boundaries.
\end{abstract}

\vspace{2pc}



\section*{Disclaimer}
\debug{This is a manuscript submitted to the journal ``Modelling and Simulation in Materials Science and Engineering''.}

\section{Introduction}
Heterogeneities of constitutive and microstructural state variables, such as stress, strain, orientation, or dislocation density, play a pivotal role during metal forming and downstream annealing treatment \cite{Kocks1975,Hosford2010,Humphreys2017}. These heterogeneities form as a consequence of local differences during dislocation self-organization \cite{Eshelby1951,Hall1951,Kocks1975,Mughrabi1983,Hirth1992} or geometrical constraints which crystalline interfaces imprint on the individual deformation mechanisms \cite{Bayerschen2016}. Especially the accumulation of misorientation and dislocation density at such interfaces \cite{Beaudoin1993,Beaudoin1996,Raabe2002,Quey2012} generates local conditions and gradients which facilitate the formation of dynamic or static recrystallization nuclei \cite{Dillamore1974a,Sakai1984,Wusatowska2002,Miura2004a,Miura2004b}. For this reason, considerable experimental effort \cite{Miller2014,Bayerschen2016} has been devoted to the characterization of deformation microstructure heterogeneities, both in 2D \cite{Sun2000,Mishra2009,Calcagnotto2010,Digioacchino2013,Wright2014,Subedi2015} and recently also 3D \cite{Pokharel2014,Pokharel2015}. As a complementary activity, it has become common practice to investigate deformation microstructure evolution with full-field 3D crystal plasticity representative volume element (RVE) computer modeling \cite{Hill1963,Drugan1996,Kanit2003} and compare explicitly such results to experiments \cite{Kords2013,Reuber2014,Wong2015,Ozturk2016,Tari2018,Miller2014,Zhao2016,Maire2017,Tutcuoglu2019,Roters2019}. 

Most commonly, such computer simulation studies report descriptive statistics of tensorial state variable values. These are either presented as histograms or exemplary as colorful renditions of the entire RVE domain or specific sections thereof. For characterizing how state variable values are spatially distributed in the RVE volume, though, one can call such characterization strategies unnecessarily qualitative if not even insufficient. Instead, one should better quantify the distance of each material point in the RVE to its closest interface and evaluate the distributions of state variable values as a function of these distances, i.e. as spatial distributions. 

We acknowledge that there are practical challenges associated with such quantification task. Especially when one aims for highest possible statistical significance and when full-field spectral method models are used \cite{Moulinec1994,Moulinec1998,Lebensohn2001,Lebensohn2012,Eisenlohr2013,Diehl2016,Diehl2017,Roters2019}: firstly, simulations for heterogeneity build-up need sufficiently more finely resolved RVE domains than are typically used for flow curve predictions. The former create higher computational costs because statistical significance demands to include not only as many grains as possible but also an as fine as possible spatial resolution for each of them. In fact, dislocation density or orientation gradients build-up in three dimensions and evolve as a function of grain orientation and neighborhood \cite{Raabe2002}. For simulations which demand fine discretization, spectral method models with periodic boundary conditions provide an advantage because of comparable lower numerical costs than similarly resolved finite element crystal plasticity models \cite{Diehl2016}. 

Nevertheless, this practical advantage comes with a second challenge: the spectral method typically uses volume sampling approaches instead of tracking explicit material points on the grain or phase boundaries. In other words, the advantage of the spectral solver for flow curve computations turns into a disadvantage because evolution of the interface geometry must be determined during post-processing. 

The requirement to analyze distances in the deformed configuration, i.e. distances as they are measured in an experiment where the sample is typically prepared after having been deformed, defines a third challenge: successive shape distortion of the RVE domain has to be accounted for. This is a particular challenge when executing above simulations with periodic boundary conditions. In such case, grains in contact with and wrapping around the RVE domain boundary should ideally get analyzed as well as to not make compromises in statistics. These grains, though, are not only frequently physically fragmented as a result of deformation but show up as eventually multiple logical fragments on opposite RVE domain wall sides. 

Another challenge to master during any quantification of spatial correlations is to collect ideally results from all material points, again to get highest statistics. For simulations with several million material points and eventually dozens of strain steps, though, this calls for efficient post-processing tools to machine off datasets in the order of possibly several hundred gigabyte volume. In other words, the post-processing tools should be parallelized. As a fifth challenge, any post-processing of state-variable-to-interface-distance correlations for spectral method models calls for grain reconstruction methods because grains accumulate internal misorientation as they get deformed. 

At least for the grain reconstruction challenge, post-processing tools were developed in the SEM/EBSD community \cite{Schwartz2009,Dillard2007,Rowenhorst2010,Konijnenberg2012,McKenna2014,Ullah2014}. Consequently, we assessed these \cite{Konijnenberg2012,Konijnenberg2013,Groeber2014,Bachmann2011} with respect to their potential for immediate application to solve our reconstruction and distancing task. We found not a single tool, though, which combined grain reconstruction functionality for arbitrarily shaped, three-dimensionally distorted surplus periodically constrained point cloud data and documented application for multi hundred million point cloud datasets.

This gap motivated our research contribution. Therein, we develop methods which cope with all above listed challenges. Specifically, we implement statistical quantification methods which characterize how state variable values are distributed as a function of material point distance to grain boundaries. We implement two grain reconstruction methods. Next, we apply these in three different distancing methods to compile a comparative assessment of long range spatial distributions for state variable values. In the case studies we focus on single-material-point-resolved quantification of the Cauchy stresses and quantify also the disorientation with respect to the mean orientation of the grain. Cauchy stress and disorientation are quantities of frequent interest and controversy with respect to their tendency to form spatial gradients at interfaces.

Furthermore, we report how such methods can be implemented into an efficient strong and weak scaling open source tool for a workstation or computer cluster. Therewith, we supplement the D\"usseldorf Advanced Material Simulation Kit (\dm \cite{Roters2019}) with a tool which encourages the crystal plasticity community to explore also their spatially resolved data in further quantitative detail.

\section{Methods}
\subsection{Defining the data analysis task}
This work presents a specific analysis pipeline to post-process \dm \cite{Roters2012,Eisenlohr2013,Diehl2016,Roters2019} simulation results. Two grain reconstruction, three material point to interface distancing methods, and additional post-processing algorithms were developed. The pipeline enables the quantification of spatial correlations between specific material point state variable values and each material points' closest projected distance to an interface. The individual steps of the pipeline are detailed in the appendix. The pipeline was implemented into a parallelized original C/C++ program. Given that the entire source code is provided as open source, we report only the key steps of the pipeline and the strategies we employed for an efficient implementation.


The pipeline answers the following data mining question: how are state variable values distributed spatially as a function of the projected distance to the nearest interface for a given collection of strain steps? We define a strain step as a state variable value dataset which is probed from an RVE crystal plasticity simulation at a specific true strain value. From now on, it is assumed that this dataset contains results for every material point which supports the RVE domain volume.

Interfaces are either grain boundaries or phase boundaries. In this paper, we focus on single phase microstructures, and thus study grain boundaries. Our methods are formulated general enough, though, to be applicable to phase boundaries, provided their geometry is implicitly spatially resolved with material points. We report our methods for three-dimensional applications, the simplification to two dimensions is straightforward.

Under above provision the input data to the pipeline consists, for each strain step, of the full set of $N$ material points. These points, identified as $p_i$, define positions $x_i$ in the deformed configuration. We define and analyze our results in the laboratory Cartesian coordinate system $\in {\mathcal{R}}^3$. As such, all local displacements of a material point versus its initial position in the unloaded case are accounted for \cite{Eisenlohr2013}. This makes the results comparable to the situation in experiments where a plastically deformed microstructure is probed typically by a posteriori sample preparation. For this coordinate system convention the initial grid of integration points becomes successively and irregularly distorted (Fig. \ref{RVESetup}). This calls for the development of point cloud processing methods which back out the geometry of each grain and consider the fact that typical \dm simulations use full three-dimensional periodic boundary conditions. Each material point $i$ has an associated set of state variable values $\{\bm{s_i}\}$. Members of this state variable set are for instance the local deformation gradient $\bm{F_i}$ or the orientation $\bm{q_i}$. Although, not all these variables are state variables in the microstructural sense, we have opted to refer to these computed quantities as state variables for simplicity.

Depending on which full-field model of the crystal plasticity community is employed, different strategies are in place to define an initial grid of integration points to represent the microstructural volume inside the RVE domain. Cubic grids are the most commonly employed strategy. This results in an implicit representation of the grain or phase boundaries. Full-field spectral method crystal plasticity simulation tools like \dm \cite{Diehl2016,Roters2019} also use such volume sampling strategy. This avoids explicit bookkeeping of the interface network geometry and thus cuts numerical costs. As a disadvantage, though, the interfaces have to be reconstructed from the point cloud data using post-processing methods, like the ones described in the following, to enable a quantification of above distance correlations. The present pipeline provides an original solution to solve this task. The pipeline is implemented in a supplemental \dm post-processing data mining tool which we call damaskpdt for short, where the acronym pdt stands for post data treatment.

\subsection{Pre-processing stage}
The processing of the strain step data is distributed across the nodes of a cluster computer. Each process handles complete strain steps (at least one). The processes independently read all material point and respective state variable values from the simulation results file. In this work the input is parsed from the traditional binary container format file. An extension of our tool to use the recently implemented HDF5 container format \cite{Diehl2017,Roters2019} remains as a straightforward task for the future.

After loading the strain step data, each process evaluates the data independently. This pre-processing stage includes the calculation of the material point positions $x_i$ in the deformed configuration and of the equivalent stress and strain tensors for each material point plus an RVE-averaging of these quantities reimplementing methods previously described \cite{Roters2012,Eisenlohr2013,Diehl2016,Roters2019}. Once accomplished, three point clouds for each strain step $n$ are defined to simplify all subsequent processing steps: the first point cloud is \mNull . It specifies the ensemble of the original $N$ material points in the deformed configuration. The second point cloud \mNulleps specifies the ensemble \mNull surplus all its 26 periodic image points inside a bounding box with thickness $\epsilon$ about the bounding box to \mNull. Here, $\epsilon$ defines a small fraction of the largest edge length of the box. \mOne , the third point cloud, specifies all members of \mNull and all their 26 periodic image points. Subsequent processing involves region queries on all three point clouds. Special data structures were used to accelerate these queries (section \ref{ImplementationTricks}).

\subsection{Grain reconstruction}
\paragraph{Remark on classical grain reconstruction with \dm}
In the past, attempts were made to reconstruct the grains via executing a clustering analyses on \mNull. Specifically, the grains were built as a cluster of material points with points within a critical distance of a few point-to-point distance units and less than an a priori defined disorientation among each other. However, such method cannot rigorously merge all periodic images of a grain. Especially not those on opposite sides of the RVE domain unless a very large critical distance is chosen. Such choice, though, will occasionally merge second- or even higher-order neighboring grains into a single object. Especially so when these neighbors have similar orientations. Therefore, in this work two different approaches are proposed to reconstruct the grains. The methods borrow conceptually from achievements made in the SEM/EBSD community \cite{Schwartz2009,Bachmann2011,Konijnenberg2012} but implement these concepts into a tool taking into account the periodic boundary conditions and showing a much better scaling in parallelization.

For any implicit representation of an interface network, here exemplified for grain boundaries, a reconstruction of the interfaces is a necessary step to enable us to build correlations between state variable values at the material points and the distances to the closest interface. Only in the special case when stress to distance correlations are sought and disorientation based distancing used, grain reconstruction is an optional step in our approach. In what follows, two methods for partitioning the material points into a polycrystalline aggregate are specified:

\paragraph{Modified grain reconstruction methods developed in this work}
The first grain reconstruction method interprets the texture ID, or texture index, initially assigned upon microstructure synthesis, as the grain ID. This has been the common strategy so far when studying in-grain orientation gradient build-up: either explicitly \cite{Beaudoin1993,Quey2012} or implicitly by reporting orientation spread via pole figures \cite{Raabe2002}. This grain reconstruction method is tagged TEX in the results section.

The second grain reconstruction method addresses the challenge that when grain fragmentation becomes significant, referring still to the initially instantiated grains is no longer correct. Then, the individual grain fragments should be distinguished. Three requirements should be fulfilled for this purpose: the method should work for irregular point cloud data; it should be able to handle full, i.e. three-dimensional periodic boundary conditions; the method should be sensitive to gradual orientation gradient build-up.

Graph clustering is one method which fulfills all these criteria. Application examples for characterizing microstructures with graph clustering methods, though, remain few. One is the Fast Multiscale Clustering (FMC) \cite{Mcmahon2013,Loeb2016} method known in the SEM/EBSD community. Other examples, on which this work settles, are community detection algorithms \cite{Newman2004,Dancette2016} that find frequent application in analyzing human communication and interaction patterns for social media platforms. Exemplarily, this work uses the community detection method employed in the study of Dancette et al. \cite{Dancette2016}. Specifically, we use the open source implementation of the Louvain community detection method by Blondel et al. \cite{Blondel2008,Blondel2015}. This grain reconstruction method is tagged LOU in the results section.

\paragraph{Extraction of a single periodic image per grain}
Once all material points are tagged with a grain ID, in principle one can compute the distance correlations. For arbitrarily deformed domains with periodic boundary conditions, though, practical challenges remain: first, macroscopic and microscopic shape distortions of the RVE domain walls have to be accounted for. Second, grains in contact with the RVE domain walls are fragmented into multiple pieces, possibly laying on opposite sides of the RVE domain. It is useful to merge these pieces into a single grain object with simpler geometry. To place this object into a cuboidal box is even better. Thereby, all downstream processing operations specific to this grain can be easier handled numerically. Consequently, a geometry simplification step is executed. Therein, first the individual periodic fragments are fused using a clustering algorithm. Thereafter, a single representative periodic image is chosen. Once chosen, all material points associated to the merged grain are packed into a grain-local bounding box. Given the a priori labeling of the material points with grain IDs, the geometry simplification step is a trivial parallel task. This allows us to employ efficient multithreading approaches developed in the grain growth modeling community to speed up the analyses \cite{Miessen2015,Miessen2016,Miessen2017}. 

In detail the simplification step works as follows: each grain $k$ is processed independently. First, all material points in \mNull are periodically replicated to define the superset \mOne . Using the grain reconstruction, points are tagged with their grain ID $k$. Subsequently, a clustering algorithm is applied to merge the individual fragments into a number of complete grain periodic images. In this study, the DBScan \cite{Ester1996} algorithm was used. It assigns all points which are connectable by spherical regions with radius $R$ to the same cluster. Specifically, a radius of $\sqrt{3}$ times the initial point-to-point distance was used. This strategy is applicable as long as no grain wraps periodically around the entire domain or, similarly, periodic fragments in \mNull on opposite sides of the RVE domain come closer to one another than the above defined critical distance. Next, one representative cluster is picked. Using this strategy the entire point cloud \mOne is constructed using multithreading. Finally, we compute the global bounding box to all grains' local bounding boxes.

\paragraph{Rediscretizing the global bounding box}
Next, we define a rediscretization of this global bounding box volume. The actual rediscretization, however, was applied to volume inside the local bounding boxes only to improve efficiency. Cubic voxels of edge length $0.5$ times the initial point-to-point distance were used. Once rediscretized, the information contained in \mOne is used to identify the nearest neighboring material point to each voxel.

\subsection{Distance quantification}
\paragraph{Disorientation based}
For this distancing method, tagged DIS in the results section, the challenge of grain reconstruction is ignored completely. Instead, position and orientation data for points in \mNulleps are used. In this work a guard zone thickness $\epsilon$ of $0.1$ times the initial RVE domain edge length was used. With these definitions the distance values are computed by finding, for all material points in \mNull, all their respective neighbors in a spherical region of radius $R = 0.1$. Again $R$ is in multiples of the initial RVE domain edge length.

Next, the identified neighbors are sorted in increasing order with respect to their Euclidean distance to the inspected material point. Finally, the distance value for the closest, above a critical threshold $\Theta_c$ disoriented pairing, is taken as the distancing result. In this work, a threshold of $\Theta_c = \SI{15}{\degree}$ was used. If no closest neighbor among the candidates was found, the material point made no contribution to the correlation statistics.

In a nutshell, the advantage of the DIS distancing method is that no grain reconstruction is required as long as no disorientation-to-distance correlations are desired. The independence of each material point is another advantage because calculations are trivial parallel. Without taking the spatial arrangement of neighboring points into account, though, a clear disadvantage of the DIS distancing method is that only the distance but not the displacement is accessible.

\paragraph{Signed distance/ voxelization based}
Enabling the identification of normal distances to the grain boundary is the motivation for and key strength of the second distancing method. It is tagged as SDF in the results section. The key idea is to combine attractive mathematical properties of signed distance level set functions with latest method developments achieved for scalable grain coarsening simulations \cite{Osher1988,Elsey2013,Scholtes2015,Miessen2017}. 

Level set function methods parameterize the position of interfaces implicitly as the iso-contour of a real-valued level set function $\Phi(x)$ to an a priori defined iso-value. A signed-distance level set function (SDF) is a particular useful definition to choose as a level set function because when making this choice surplus defining the grain boundary contour for each grain as $\Gamma := \{ x \in \Omega \mid \Phi(x) = 0 \}$ with $\Gamma \in \Omega \in {\mathcal{R}}^3$, a distance level function defines the (normal) distance of continuum points to the the grain boundary contour. This choice, i.e. $\Phi(x) \in \Omega \in {\mathcal{R}}^3$, has the key advantage that $\Vert \nabla \Phi(x) \Vert = 1$ holds for most points $\in {\mathcal{R}}^3$ one practically encounters inside and outside the contour. Consequently, a consistent outer unit normal vector can be computed to each point on the contour by evaluating $\bm{n(x)} = \frac{\nabla\Phi(x)}{\Vert \nabla \Phi(x) \Vert} = \frac{\nabla\Phi(x)}{1}$ numerically. In effect, this allows us to calculate arbitrarily projected distances to the contour. In other words the SDF provides access to a distance vector.

A two step procedure was employed to compute a signed distance function for each grain. The procedure builds on the previously defined discretization of the bounding boxes about each merged grain. First, $\Phi(x)$ was initialized to positive values ($+h$) for all voxel assigned to the grain under inspection and negative values ($-h$) for points outside. The scalar $h$ was set equal to the cell width used for the global rediscretization of the domain. In a second step, the distance information was propagated with the fast sweeping method \cite{Zhao2004,Miessen2017}.


\paragraph{Tessellation based}
Also the third method, tagged VOR in the results section, delivers vectorial distances. Furthermore, it explicitly reconstructs a contour hull for each grain. The key idea of the method is to build each grain volumetrically from a collection of Voronoi polyhedra. These can be directly obtained from a Voronoi tessellation of the respective material point cloud for each grain. Formally, the tessellation based method is based on the approach detailed by Bachmann et al. \cite{Bachmann2011}. We introduce two improvements, though: first, we generalize the method for applications on distorted point clouds with periodic boundary conditions. Second, we introduce a hybrid parallelized solution which executes substantially faster and might provide an avenue to explore for MTEX as well. 

The parallelization alleviates much of the higher numerical costs, which the VOR method has compared to the DIS or SDF methods. It is not necessary within VOR to rediscretize the bounding boxes. Instead, we directly operate on the specific grain-local subset of material points from \mOne, specifically those points inside the bounding box to each grain $k$.

\begin{figure}[!ht]
\centering
	\subfloat[a][Distancing contours with mixed curvature signs]{\includegraphics[width=0.45\textwidth]{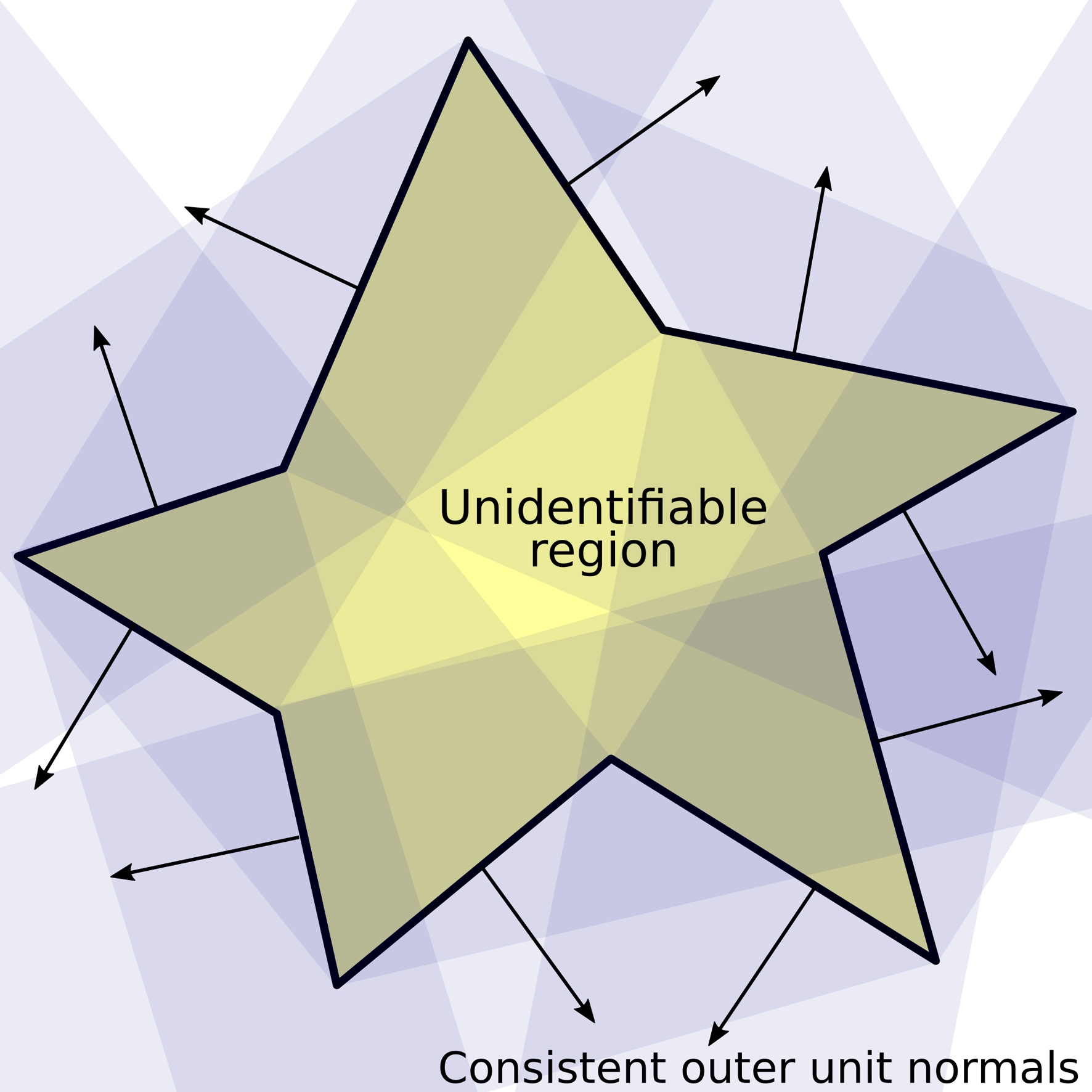}\label{VORStar}}
	\quad
	\subfloat[b][Exemplar grain contour hull]{\includegraphics[width=0.45\textwidth]{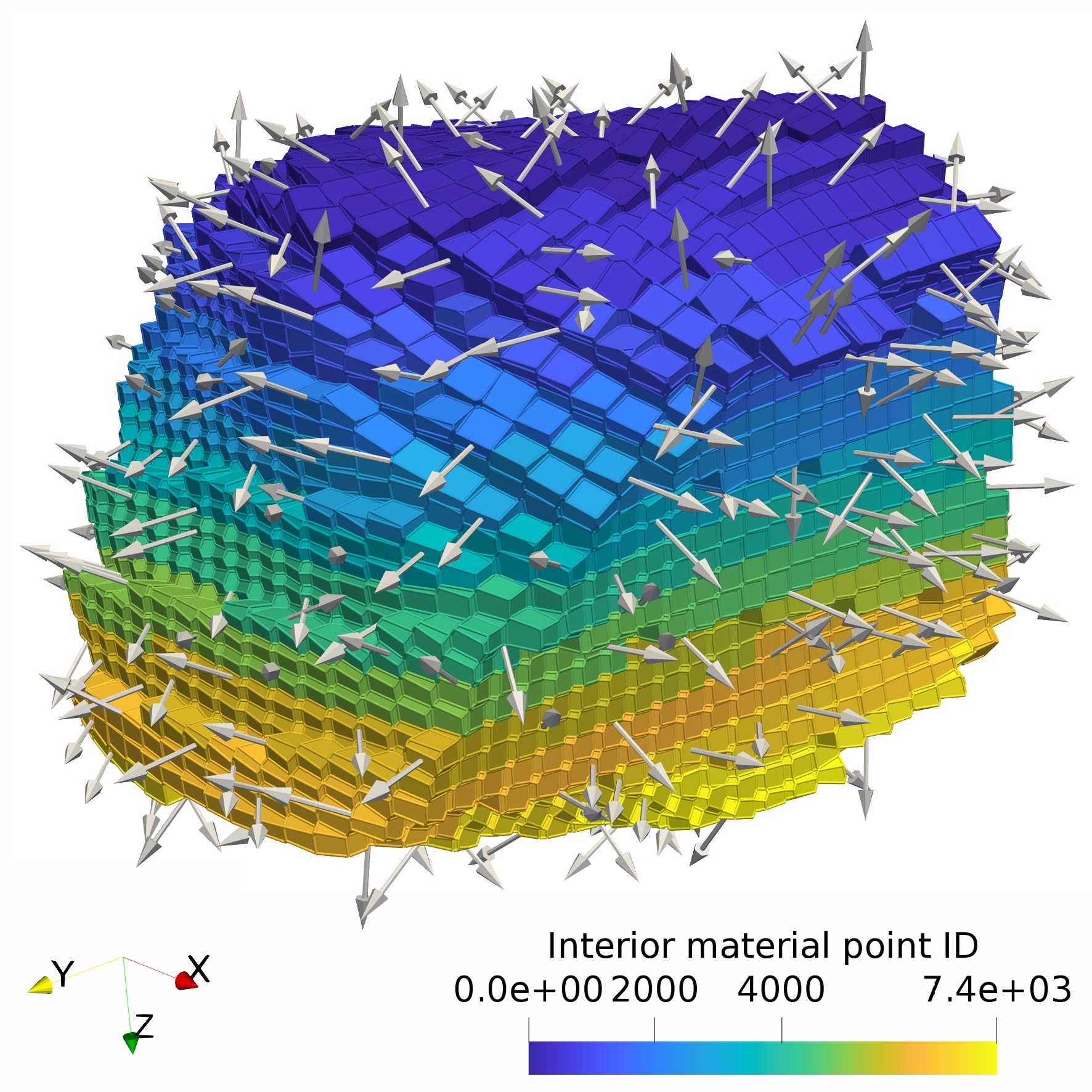}\label{VORVis}}
\caption{a) Compositing Voronoi cells into a larger polyhedron can result in a polyhedron with concave and convex sections. This demands additional care when computing projected distances to the contour hull facets. b) An exemplar result of the tessellation based contour hull extraction algorithm. A random subset of outer unit normal vectors is shown.}
\label{VORMethods}
\end{figure}

In a first step, this local point cloud is extracted and each Voronoi cell computed. We tag each Voronoi cell with the grain ID $k$ associated to the material point. Next, we identify which of the cell facets have first-order neighboring Voronoi cells with a different associated grain ID than the currently processed grain tag. Such difference is a signature that the currently processed Voronoi cell is in contact with the grain boundary. The resulting collection of all such facets of Voronoi cells about points tagged with the grain ID under inspection is used to construct the shape of the exterior contour hull (Fig. \ref{VORVis}). The contour hull extraction algorithm is executed in parallel for each grain $k$. 

We used the \voroxx tessellation library \cite{Rycroft2009} to process the individual grain-local tessellations. Specifically, we re-implemented the library wrapping approach of Peterka et al. \cite{Peterka2014} around calls to this library. To avoid truncating Voronoi cells of the inspected grain, we enlarged the individual grain-local bounding boxes using a guard zone of three times the initial point-to-point distance. Furthermore, possible Voronoi cell contact with the local bounding box domain walls was detected using \voroxx library functionalities. Thereby, we cured any possibly incorrectly computed Voronoi cell, eventually through fattening the guard zone.

Once the exterior contour hull has been defined, the material point to contour hull distances can be computed. Given that projections on the polyhedron facets can be used, the VOR methods computes projected distances. The implementation is tricky, though, in detail because consistence and speed are desired:

\paragraph{Algorithm to compute vectorial distances within VOR}
First, a consistent outer unit normal vector for each contour hull facet is computed using the edge circulation conventions of the \voroxx library. Next, the normal and geometry of the facet is used to compute a consistent right-handed local Cartesian coordinate system for each contour hull facet. Such coordinate system can be computed because each Voronoi cell is a convex polyhedron \cite{Okabe2000}. By virtue of construction each facet is a convex polygon in 3D space. With the aid of the local coordinate system, though, the geometric description can be reduced to two dimensions. This accelerates point to polygon inclusion tests.

It remains to compute the actual distances. First, the general approach used is explained. Second, we detail how the processing of the facets has been accelerated using bounded volume hierarchy querying algorithms.

The actual distances are Voronoi cell interior-point-to-exterior-contour-hull-projected point distances. Distances were evaluated for each interior point. It suffices to sketch the procedure for a single interior point. The main task is to probe which facet of the contour hull lays closest to the point. Given that the contour hull was constructed from Voronoi cells, each facet is a polygon. 

While probing the facets for the absolute shortest projected distance we proceed as follows: firstly, each interior point gets normal projected onto the facet plane.
Secondly, we probe whether the projected point lays inside the facet polygon on this plane or outside. If the point is enclosed or on the boundary, the point is considered inside. In this case the distance value is eventually used to update the currently closest distance. 

If, though, the projected point is outside the polygon, we proceed by normal projecting the currently tested interior point on each edge segment of the polygon. If such a projected point lays on the edge, the computed distance eventually updates the current closest distance. 

If all these projections, though, do not map on a polygon edge segment, we evaluate the Euclidean distance of the currently tested interior point to each vertex of the polygon. Eventually thereby also the current closest distance gets updated. Once finished with the vertices, we proceed with the next facet of the contour hull until no further facets remain for processing.

To the best of our knowledge advanced procedures like above are necessary to handle cases where, due to the possible non-convexity of the contour hull, a consistent normal distance is not immediately defined. In fact, such cases are frequently possible. One particular simple example is illustrated with the grayed-out region in Fig. \ref{VORStar}. Evidently, there is not necessarily a projected point on a contour facet for every position inside an arbitrarily shaped non-convex polygon (here for the brighest yellowish region). At least not if one projects exclusively parallel to the direction of the facets' outer unit normals.

While getting access to a defined contour hull for each grain surplus projected distances is a clear advantage, the VOR method has also disadvantages: most importantly, the method is computationally more costly. More subtle, the edge-on compositing of a contour hull from a collection of polygons faces discontinuities of the curvature at the polygon edges. These can be cured, though, with state of the art SEM/EBSD grain reconstruction software \cite{Konijnenberg2012,Konijnenberg2013} which is another possible avenue to explore and use for reconstruction of the interface network in \dm in the future.


\paragraph{A strategy for accelerating the distance computations}
Without additional tricks, the algorithm defines a quadratic time costly computational geometry task -- for each interior point each contour facet needs inspection. Therefore, a pruning strategy was designed to reduce the total number of facets to test. The key idea is to use the fact that while we iteratively narrow down the search for the absolutely closest distance also the total number of facets which could at all provide even shorter distances reduces. Therefore, a bounded volume hierarchy (BVH) was built from the contour hull facets \cite{Bentley1975,Brinkhoff1996,Balasubramanian2012,Hedges2017}. This utility data structure organizes the locations of the facet polygons into spatial regions, such that with narrowing down our search fewer and fewer candidates need testing. The querying structure was used during above distance computations as follows: first, an arbitrary facet was probed. Next, the resulting distance value was used to query the BVH to narrow down which candidate facets to probe next. Thereafter, the algorithm works iteratively: once the current distance value gets lower than $0.9$ times the value after which the last candidate list update was queried, the candidate list gets updated. This procedure is repeated until the absolute shortest distance per interior point is found.


\paragraph{Distilling spatial distributions of disorientation and stress}
The results of each damaskpdt post-processing tool run are two ensembles of Cauchy stress tensor/distance and orientation quaternion/distance value pairs per material point. These data were finally processed into spatial distributions of stress and disorientation. Scripts for \matlab (v2017a) and the MTEX \cite{Hielscher2008,Mainprice2014} texture toolbox (v5.0.3) were used.

A two stepped protocol was executed to quantify the disorientation of each material point to the respective mean orientation of the reconstructed grain: first, a mean orientation was computed for each grain. For this task documented methods for inferential quaternion statistics \cite{Bachmann2010} were used. Secondly, these mean orientations were evaluated against the individual orientation of each material point.

The so generated value/distance pairs were binned into distance classes on the interval $[\SI{0.0}{}, \SI{24.0}{}]$ using $\SI{0.2}{}$ step. Length units are reported in multiples of the initial $[100]$-direction point-to-point distance between neighboring material points. Quantile values of the resulting sub-distributions were quantified with \matlab. Grain size distributions report how many material points are assigned per grain.

\subsection{Parallel implementation and orchestrated hierarchical data placement}
\label{ImplementationTricks}
Above pipeline was implemented into a hybrid parallelized tool. Inherent parallelism in the data mining task was exploited: specifically, each strain step is independent. We distribute the processing of the strain step ensemble using process data parallelism. Specifically, calls to the Message Passing Interface (MPI) library \cite{Gropp1999a,Gropp1999b} were implemented. Strain steps were distributed across the MPI processes in round-robin fashion. In addition, the processing of each strain step was accelerated via Open Multi-Processing (OpenMP) \cite{Chapman2007,OpenMP2019,Hennessy2012,Rauber2013,Reinders2014,Jeffers2015,Miessen2017}. Instead of using the commonly employed referencing of data items via global arrays, we actively enforce a partitioning of the material point positions and state variable values. Specifically, we split all derived quantities into thread-local data chunks. These chunks are placed preferentially in memory locations with fast connection to the execution core. This has two advantages: first, it places grain-local data closer in memory which facilitates faster reading from memory. Second, the partitioning reduces cache coherency traffic. We admit that our implementation currently uses a static load partitioning scheme. A possible disadvantage to investigate in the future is that load imbalances may be stronger. 

To efficiently execute all linear algebra operations of the pipeline, fast Fourier transformations (FFT), singular value decomposition, and eigenvalue decomposition functionalities of the Intel Math Kernel Library (IMKL) were used.

\begin{figure}[!ht]
\centering
	\subfloat[a][Hierarchical data partitioning and mapping approach]{\includegraphics[width=0.45\textwidth]{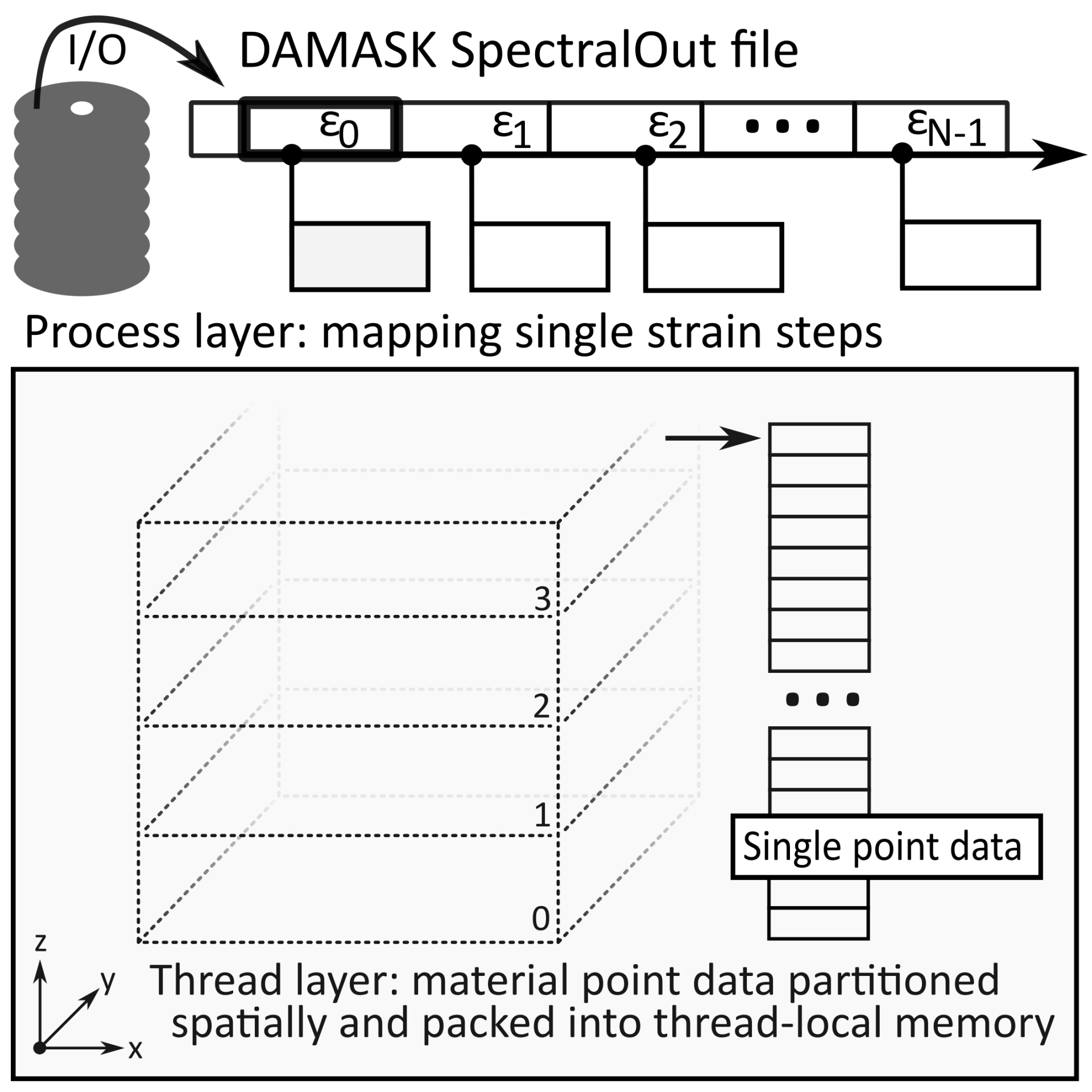}\label{PartitioningVis}}
	\quad
	\subfloat[b][Bounding box ensemble]{\includegraphics[width=0.45\textwidth]{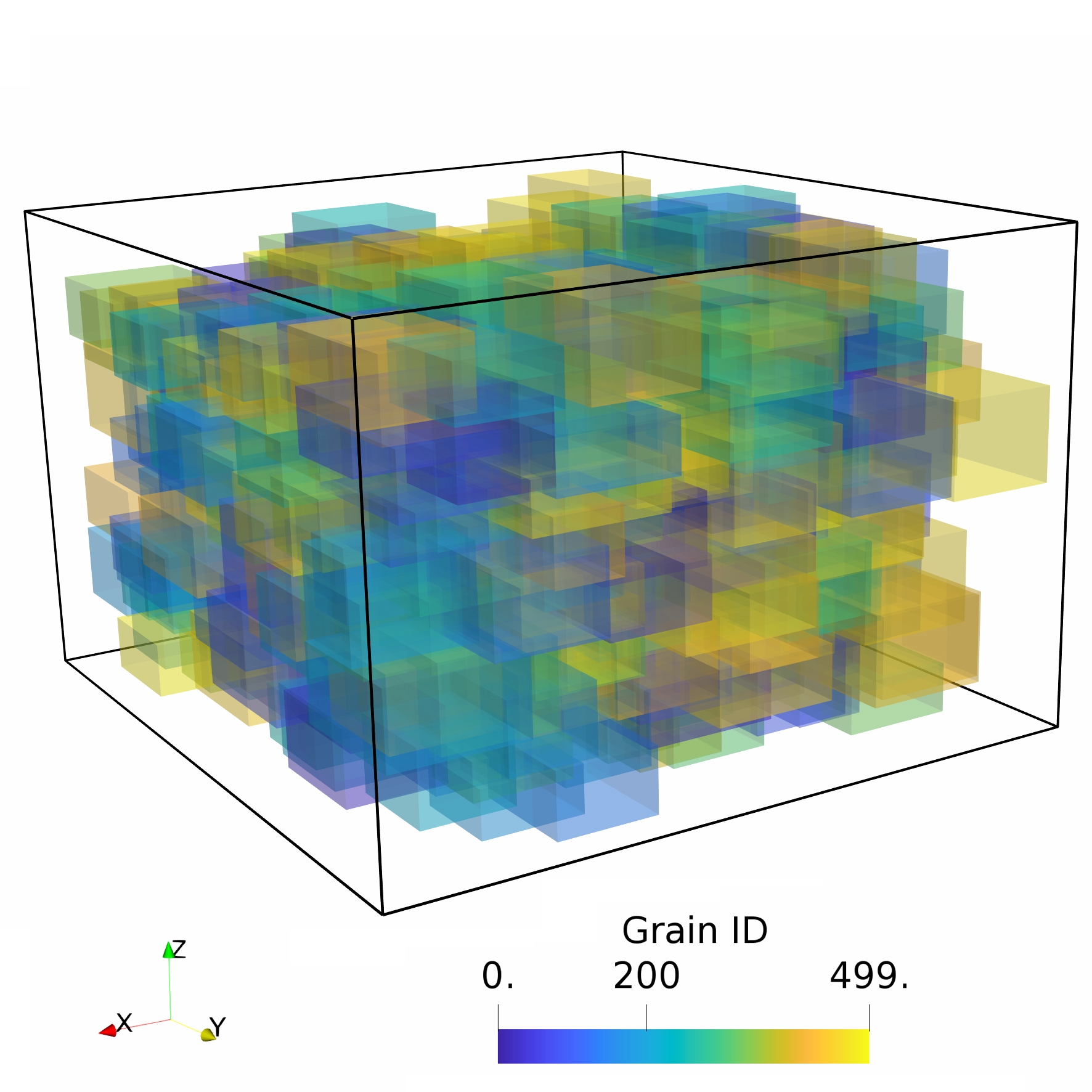}\label{BoundingBoxEnsembleVis}}
\caption{a) The processing of the strain step ensemble is partitioned hierarchically: processes are assigned always complete strain steps. These are machined off using OpenMP multithreaded data parallelism. Each process partitions the heavy data further to allow for a mapping of thread-local data into specific memory sections to improve spatial and temporal memory locality.}
\label{MethodsDataPartitioning}
\end{figure}

The subsequent grain segmentation and  characterization of the spatial descriptive statistics involves range querying tasks. Specifically, they demand the identification of all neighboring points inside a spherical volume of radius $R$. To accelerate these queries, we employed 3D box binning of \mNulleps, and \mOne . Cubic buckets of width $R$ were used. This allows us to execute constant ($\mathcal{O}(1)$) computational time complex queries. A disadvantage of this technique compared to other established fast position querying techniques \cite{Bentley1975,Patwary2015,Patwary2016} is the larger memory footprint. This detail remains as future work for a possible code optimization to reduce memory consumption.

\subsection{Verification and validation simulation setup}
\paragraph{Microstructure instantiation}
We applied our method for studying the evolution of the spatial distribution of stresses and disorientation in full-field crystal plasticity simulations. Simulations were executed with the \dm spectral solver \cite{Roters2012,Diehl2016,Roters2019}. We used the \dm Poisson-Voronoi tessellation microstructure synthesis and orientation sampling routines. Specifically, a three-dimensional single phase cubic face centered polycrystalline representative volume element (RVE) domain was instantiated. Full three-dimensional periodic boundary conditions were used. The RVE domain contained \SI{500}{} grains. These were discretized by $256^3$ grid points, i.e. material points. In effect, each grain has an average spherical equivalent radius of $20$ material points. Grain orientations were sampled randomly and assigned in a spatially uncorrelated manner.

\paragraph{Constitutive model}
We used a phenomenological crystal plasticity model \cite{Hutchinson1976}. The tensor of elastic constants was parameterized based on experimental data for Aluminium \cite{Ma2006}. The tensor components were $C_{11} = \SI{106.75}{\giga\pascal}$, $C_{12} = \SI{60.41}{\giga\pascal}$, and $C_{44} = \SI{28.34}{\giga\pascal}$, respectively. Dislocation slip on the 12 primary slip systems was assumed as the exclusive deformation mechanism. The phenomenological model was parameterized with a reference strain rate of $\dot{\gamma_0} = \SI{0.001}{\per\second}$ and a stress exponent of $n_{slip} = 20$. Stress parameters of this constitutive model were set to $\tau_0 = \SI{31}{\mega\pascal}$, $\tau_{sat} = \SI{63}{\mega\pascal}$, the hardening parameters were assumed as $a_{slip} = \SI{2.25}{}$ and $h_0 = \SI{75}{\mega\pascal}$, respectively. Isothermal conditions and initially stress free grains were assumed.

\paragraph{Load case}
The RVE was subjected to uni-axial compression along the z-axis, simulated by a deformation gradient rate $\dot{\overline{\bm{F}}}$ (Eq. \ref{DefGradStressBC}). First Piola-Kirchhoff $\boldmath\overline{\bm{P}}$ stress boundary conditions (Eq. \ref{DefGradStressBC}) were set. Stars identify unconstrained values.

\begin{equation}
\dot{\overline{\bm{F}}} = 
\left(
\begin{array}{ccc}
	*		&	0		&	0 \\
	0		&	*		&	0 \\
	0		&	0		&	-0.1 \\
\end{array}
\right)
\SI{}{\per\second}
\quad
\quad
\quad
\overline{\bm{P}} = 
\left(
\begin{array}{ccc}
	0		&	*		&	* \\
	*		&	0		&	* \\
	*		&	*		&	* \\
\end{array}
\right)
\SI{}{\pascal}
\label{DefGradStressBC}
\end{equation}

\begin{figure}[!ht]
\centering
	\subfloat[a][Deformed RVE]{\includegraphics[width=0.45\textwidth]{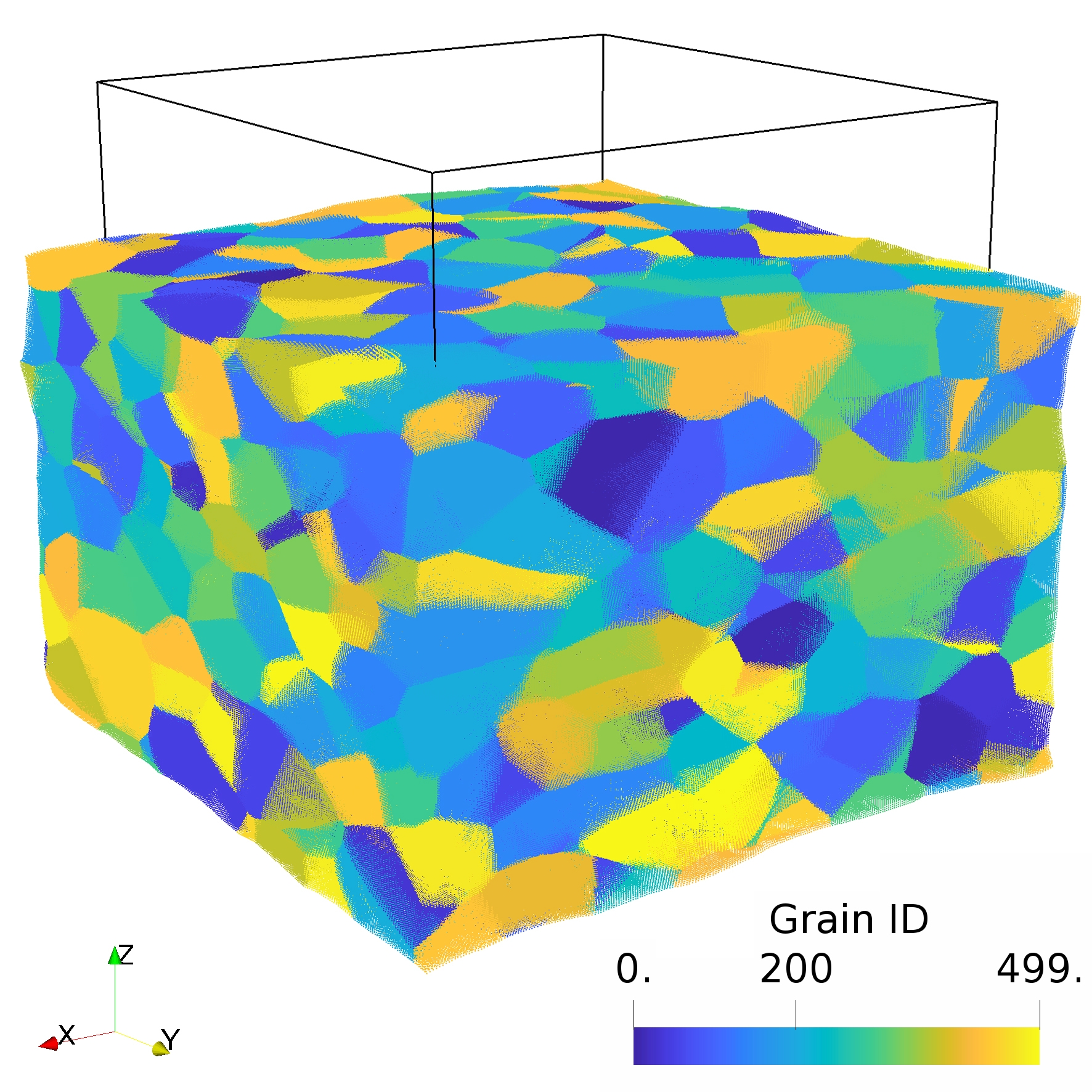}\label{RVESetup}}
	\quad
	\subfloat[b][Flow curve prediction with strain steps]{\includegraphics[width=0.45\textwidth]{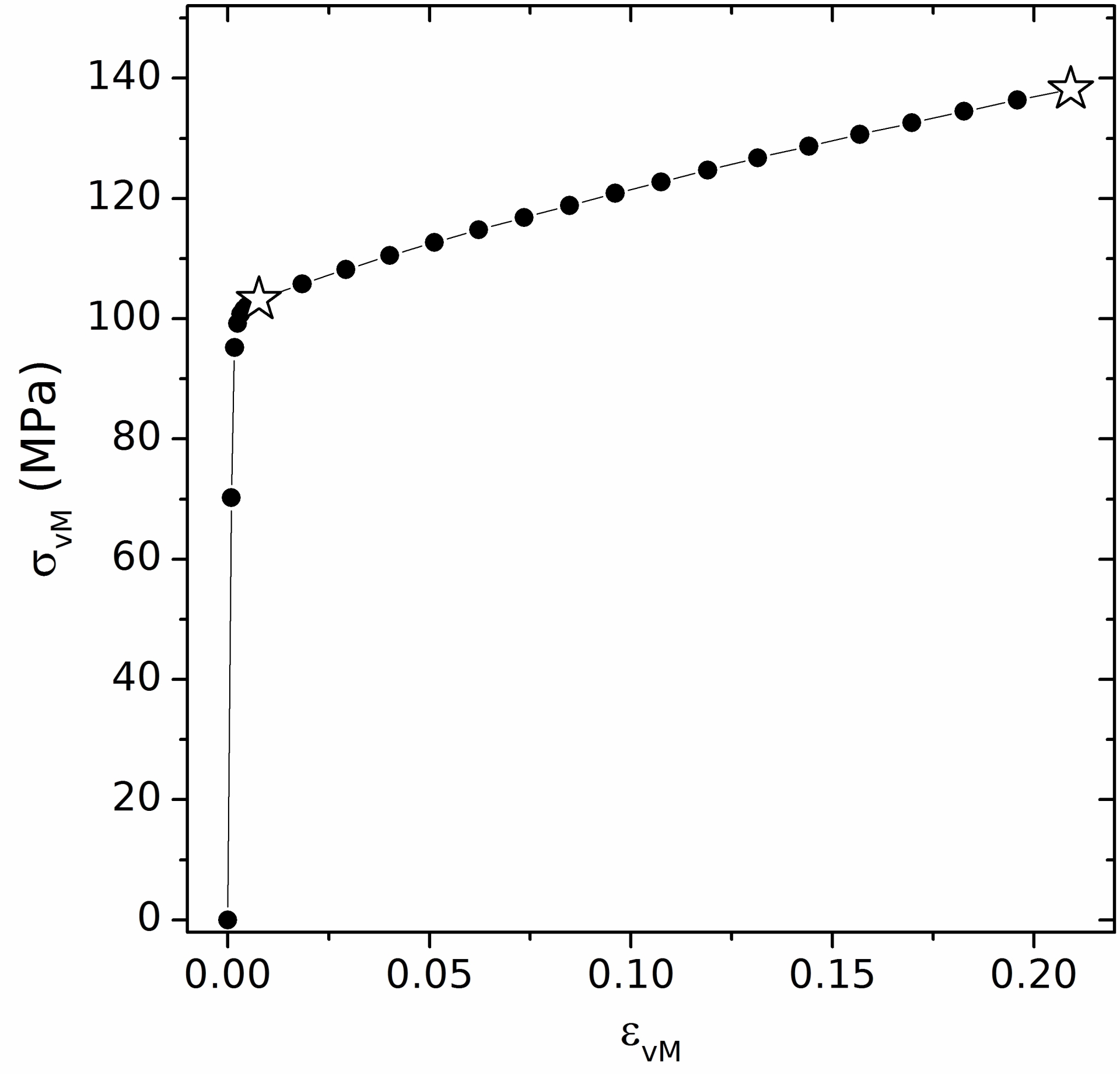}\label{Flowcurve}}
\caption{The methods were applied to a full-field simulation of a ${256}^3$ material point cube representing a polycrystal with $500$ grains. State variable values for every material point were monitored at specific total strain values as shown in b). Strain steps taken at yield (\ervat{0.01}) and incipient fragmentation (\ervat{0.21}) (indicated by the stars) are discussed in detail.}
\label{ResOverviewAllStrained}
\end{figure}

The resulting deformed RVE domain is visualized in Fig. \ref{RVESetup} exemplified for the deformed configuration in the lab coordinate system at a final equivalent von Mises strain \ervat{0.21}.

\paragraph{Simulation settings and monitoring}
Default settings of the \dm spectral solver were used. At least $4$ and at most $40$ iterations per strain step were allowed. The simulation was executed on an in-house workstation (Tab. \ref{HardwareDetails}) using it exclusively and employing one MPI process with $36$ OpenMP threads without pinning the threads. Microstructure snapshots were written at 28 selected strain values which the black dots in Fig. \ref{Flowcurve} detail. The following quantities were logged for each material point $i$ at each strain step point: the volume, the deformation gradient $\bm{F}$, its plastic $\bm{F_p}$ and elastic part $\bm{F_e}$, the first Piola-Kirchhoff stress tensor $\bm{P}$, orientation quaternion $q$, and a reference ID, the so-called texture ID ${\mathcal{T}}_i$. The latter integer specifies the discrete orientation ID which we assigned initially to each material point. The resulting binary spectralOut file occupied \SI{162}{\giga\byte} disk space.

\paragraph{Simulation data post-processing}
Post-processing was executed on TALOS, a cluster computer (Tab. \ref{HardwareDetails}). Different levels of hybrid, i.e. process and thread parallelization were benchmarked. Computing nodes were used exclusively and threads pinned (OMP\_PLACES=cores). Explicit calls to the MPI\_Wtime and omp\_get\_wtime functions were used to monitor how much time the individual pipeline steps took. I/O and non-I/O operations were distinguished. The main virtual and resident set size memory consumption was probed at the node level by parsing on the fly from the /proc/self/stat system file.

The strong multithreading speed of the tool was benchmarked with runs on a single TALOS node. Specifically, quasi-sequential runs with one process spawning one thread were compared to repetitive runs of the same study with one process spawning 40 threads. The multi node weak scalability was probed by comparing the single node results with runs employing 28 MPI processes each of which spawning 40 threads. I/O operations were performed using a GPFS parallel file system built on top of 12 logical disks. Files were striped across a RAID6 array with 10 disks each (in 8+P+Q \cite{IBM2019} configuration). The low level stripe size was \SI{4}{\mega\byte}. Supplemental MTEX post-processing was performed sequentially using a desktop PC. Given the processing consumed only a few core hours in total, these analyses were not benchmarked.

\subsection{Software details}
\dm (git commit ID: v2.0.1-992-g20d8133) was compiled with GNU v8.2 using O2 optimization. The program was linked against the PETSc v3.11.0 numerical and the MPICH v3.3 MPI libraries. The damaskpdt post-processing tool (git commit ID: caea9c52848809e0635ba7d0afb313a592e3d0dd) was compiled with the Intel Compiler and Performance Suite (v2018.4) using O3 Skylake optimization. The program was linked against the corresponding Intel Math Kernel Library (IMKL) and Intel MPI versions. Additionally, Boost (v1.66) \cite{Schling2011} and the Voro++ (v0.4.6) \cite{Rycroft2009} libraries were linked to the tool.

\begin{center}
\begin{table}[!ht]
\caption{Technical details of the computing systems used. C/S means hyperthreading core pairs per socket. Mem stands for main memory in \SI[mode=text]{}{\giga\byte}.} 
\centering
\begin{tabular}{lllll}
	\addlinespace[0.2em]
	\toprule
    \bf{System}	&	\bf{CPU}		&  \bf{C/S}    & \bf{Mem} & \bf{Operating system}	\\ \midrule
	Workstation	&	Xeon Gold 6150	&	18/2   & 576 & Ubuntu 18.04.2 LTS \\
	Cluster     &   Xeon Gold 6138  &   20/2   & 188 & SUSE Linux Enterprise Server 15 \\ \bottomrule
	\addlinespace[0.2em]
\end{tabular}
\label{HardwareDetails}
\end{table}
\end{center}

\section{Results}
\subsection{Quantifying the spatial distribution of stresses towards grain boundaries}
\begin{figure}[!ht]
\centering
   	\subfloat[a][Distance distributions]{\includegraphics[width=0.45\textwidth]{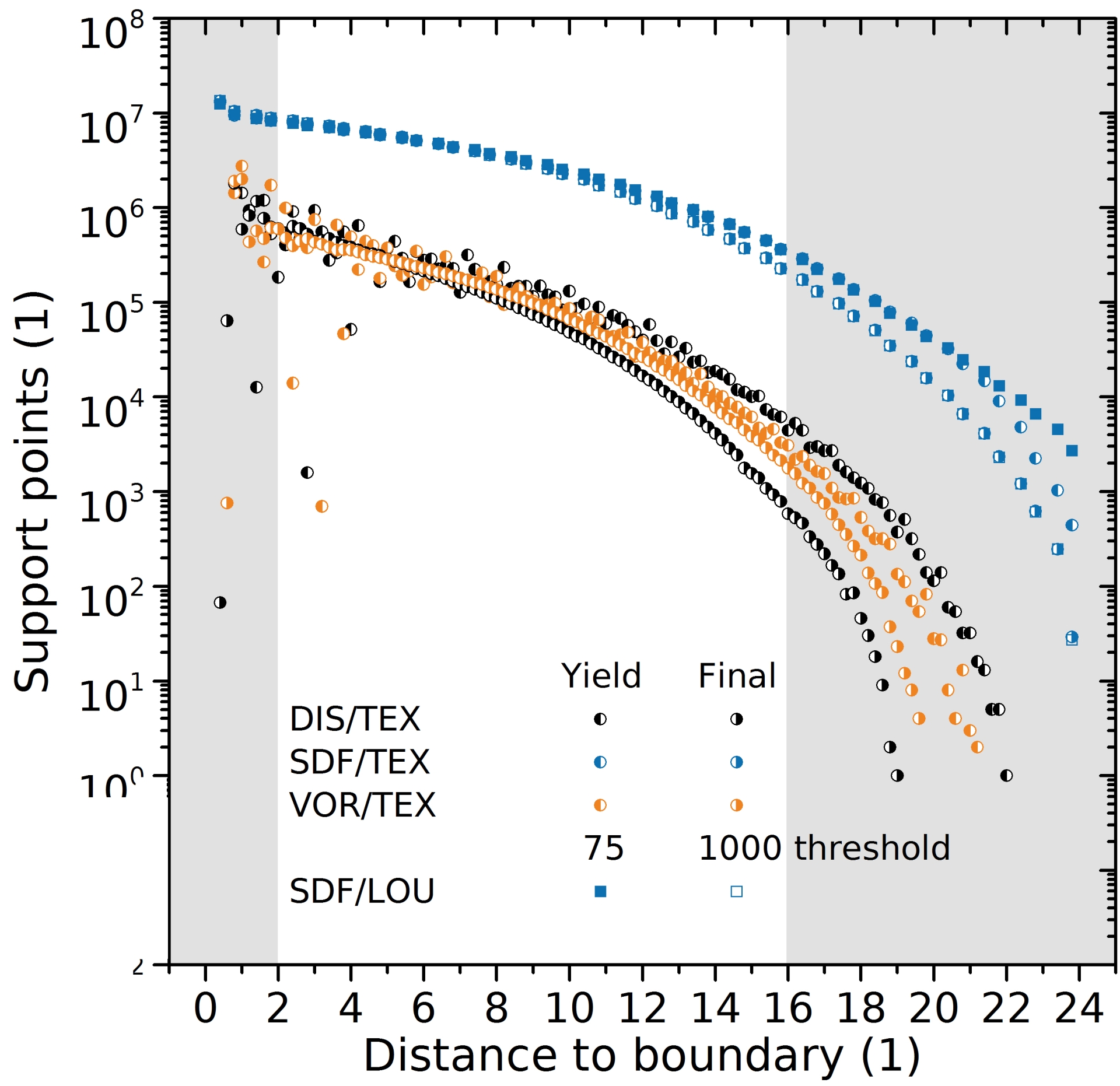}\label{DistancesALL}} 
   	\quad
    \subfloat[b][Texture ID based]{\includegraphics[width=0.45\textwidth]{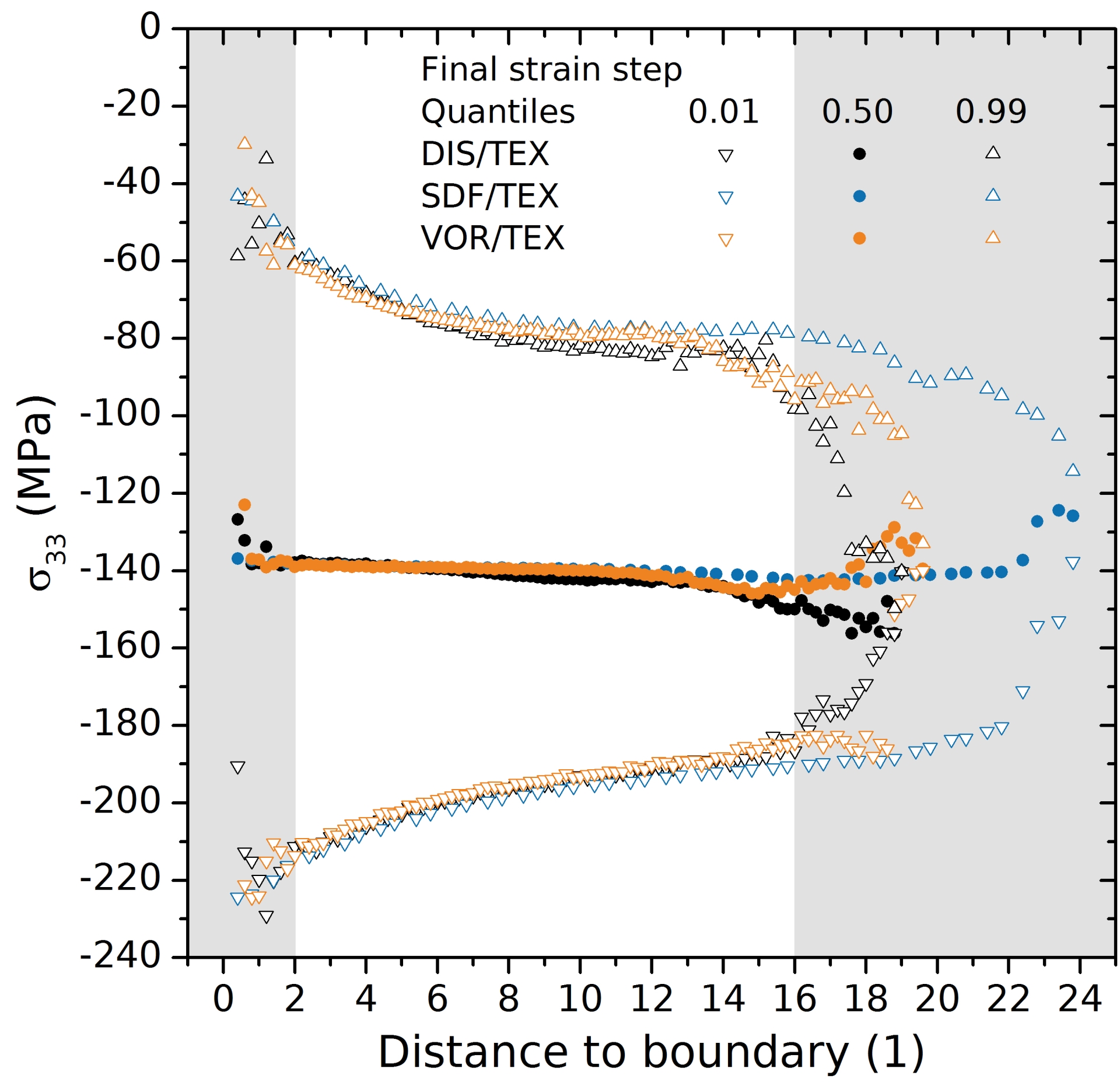}\label{StressALLTEX}} 
	\quad
	\subfloat[c][Louvain based SP]{\includegraphics[width=0.45\textwidth]{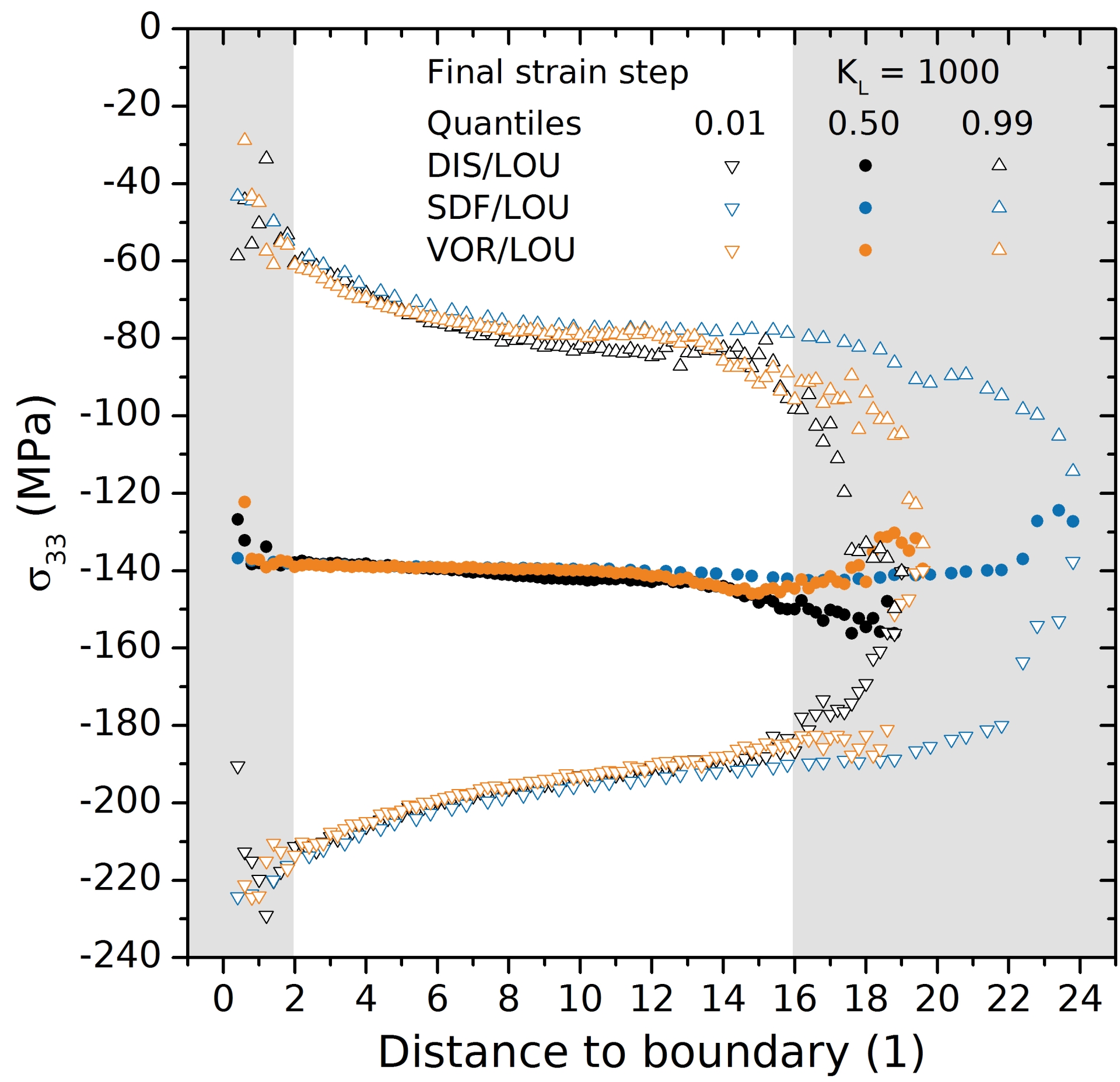}\label{StressALLLOU}}
	\quad
	\subfloat[d][Louvain based WP]{\includegraphics[width=0.45\textwidth]{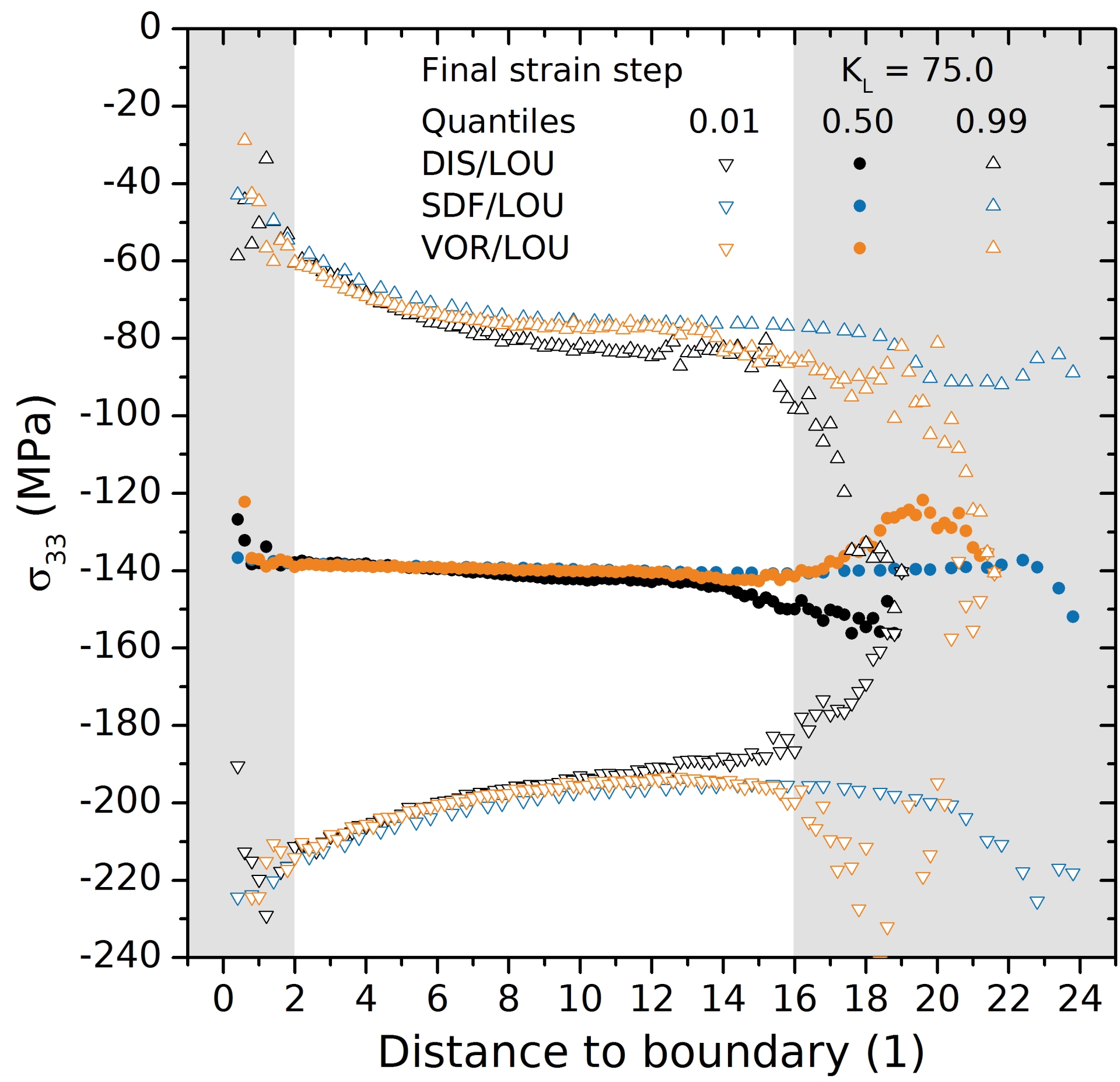}\label{StressALLLOU75}}
\caption{Cauchy stress spatial distributions as a function of the distance to the nearest boundary, here exemplified for the largest stress component $\sigma_{33}$. Abbreviations discern the distancing methods: disorientation (DIS), signed distance / voxelization (SDF), or tessellation (VOR) based, the grain reconstruction method (texture index TEX or Louvain LOU based), and whether large disorientations within a cluster were strongly (SP, $k = 1000$) or weakly (WP, $k = 75$) penalized during grain reconstruction Grayed-out regions detail where numerical effects of the spectral method are expected strongest --- either due to numerical effects or finite counting.}
\label{StressDistrosAssessment}
\end{figure}

The first application of above methods is the statistical quantification of how stresses are distributed in the RVE volume with respect to the distance of each material point to the nearest grain boundary. Using the two developed grain reconstruction methods in combination with the three distancing methods allowed for rigorous and comparative analyses on the same dataset. The results are summarized in Fig. \ref{StressDistrosAssessment}. They document the statistics at final strain when the stress distributions are broadest and highest in terms of absolute values. All distances are reported in multiples of the unit distance between two \dm material points with respect to the unloaded start configuration.

Specifically, all methods identify with very similar significance and quantile values that the average $\sigma_{33}$ compressive Cauchy stress is \SI{140}{\mega\pascal}. Only an at most \SI{10}{\percent} absolute variation from this average value is observed when probing deeper into the grain interior. Distances larger than approximately \SI{15}{} distance units from the boundary should not be interpreted because the corresponding numerical support is finite counting limited, i.e. only very few points belong to this group due to an average grain radius of \SI{20}{} distance units. 

Comparing the distributions for each distance class and method combination (Fig. \ref{DistancesALL}) conveys that all methods capture the flattening of the grains in the compression direction. Consequently, higher counts for the same distance class are observed for the strain step at yield versus the one at final strain.

The results for the lower distance classes document a rigorous quantification of the numerical effects inherent to spectral based methods in general, and those specific for the \dm spectral solver: stresses at material points in the vicinity of numerical discontinuities, here the grain boundaries, show consistently different mean stresses (Figs. \ref{StressALLTEX}/\ref{StressALLLOU}) (\SIrange{10}{14}{\percent}). Finding this difference consistently for material points at the discontinuities suggests that this is attributable to the Gibbs phenomenon \cite{Gottlieb1997,Gelb2007}. Latter is known as a systematic intensity overshooting in frequency space when attempting to Fourier expand at discontinuities.


\subsection{Quantifying grain orientation spread accumulation towards boundaries}
Motivation for this second application exercise comes from frequent literature findings which reported that material volume in the vicinity of grain \cite{Ashby1970} and phase boundaries \cite{Habiby1993} is differently disoriented than volume in the grain interior. Possible presence of such localized volume with different orientation has been identified as one reason why static and dynamic recrystallization nuclei \cite{Bellier1977} nucleate frequently on the faces or junctions of the grain boundary network \cite{Miura2004a,Miura2004b,Miura2005}.

\begin{figure}[!ht]
\centering
	\subfloat[a][Texture ID based]{\includegraphics[width=0.45\textwidth]{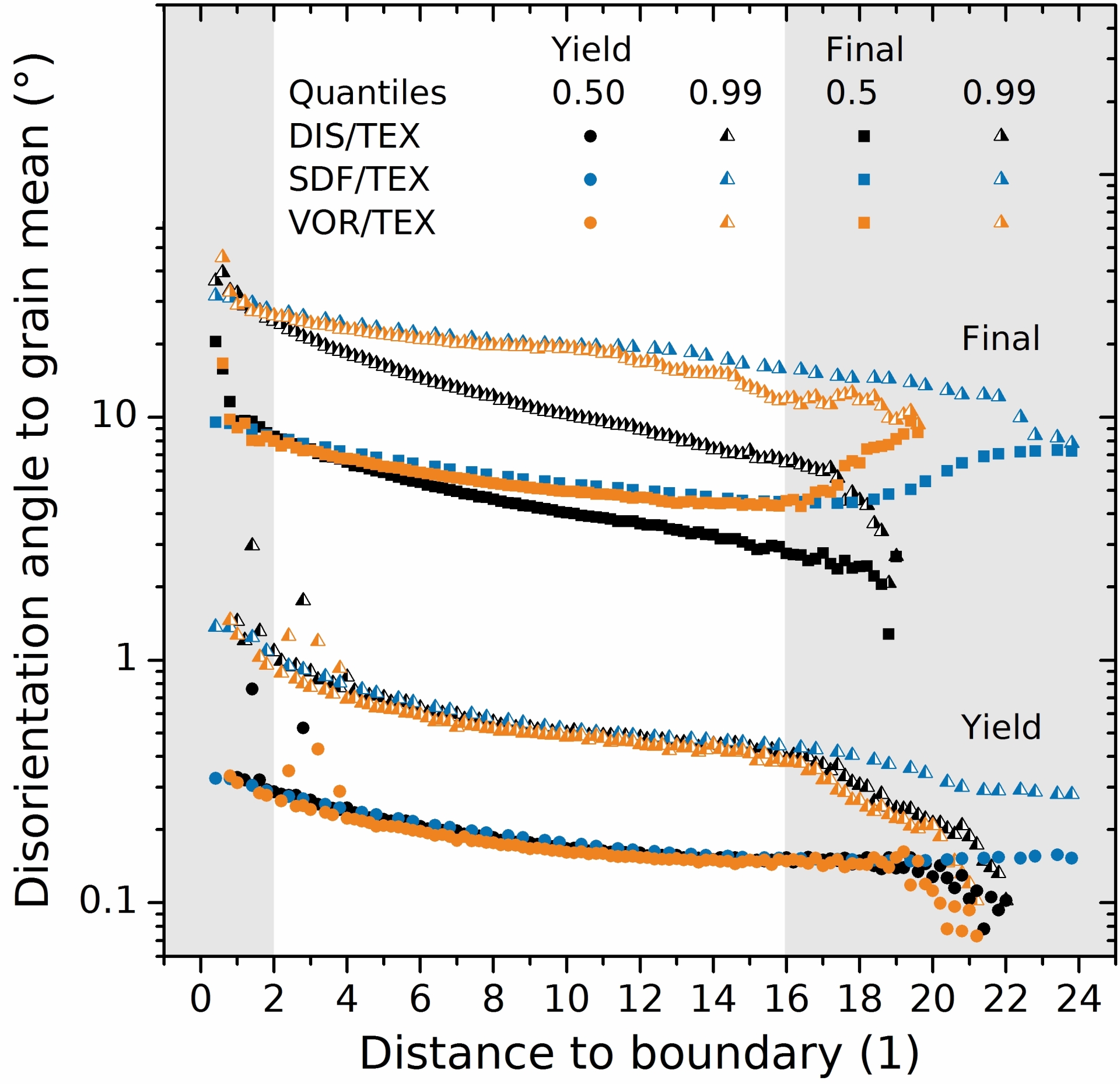}\label{DisoriDistroALLTEX}}
	\quad
	\subfloat[b][Louvain based WP vs SP]{\includegraphics[width=0.45\textwidth]{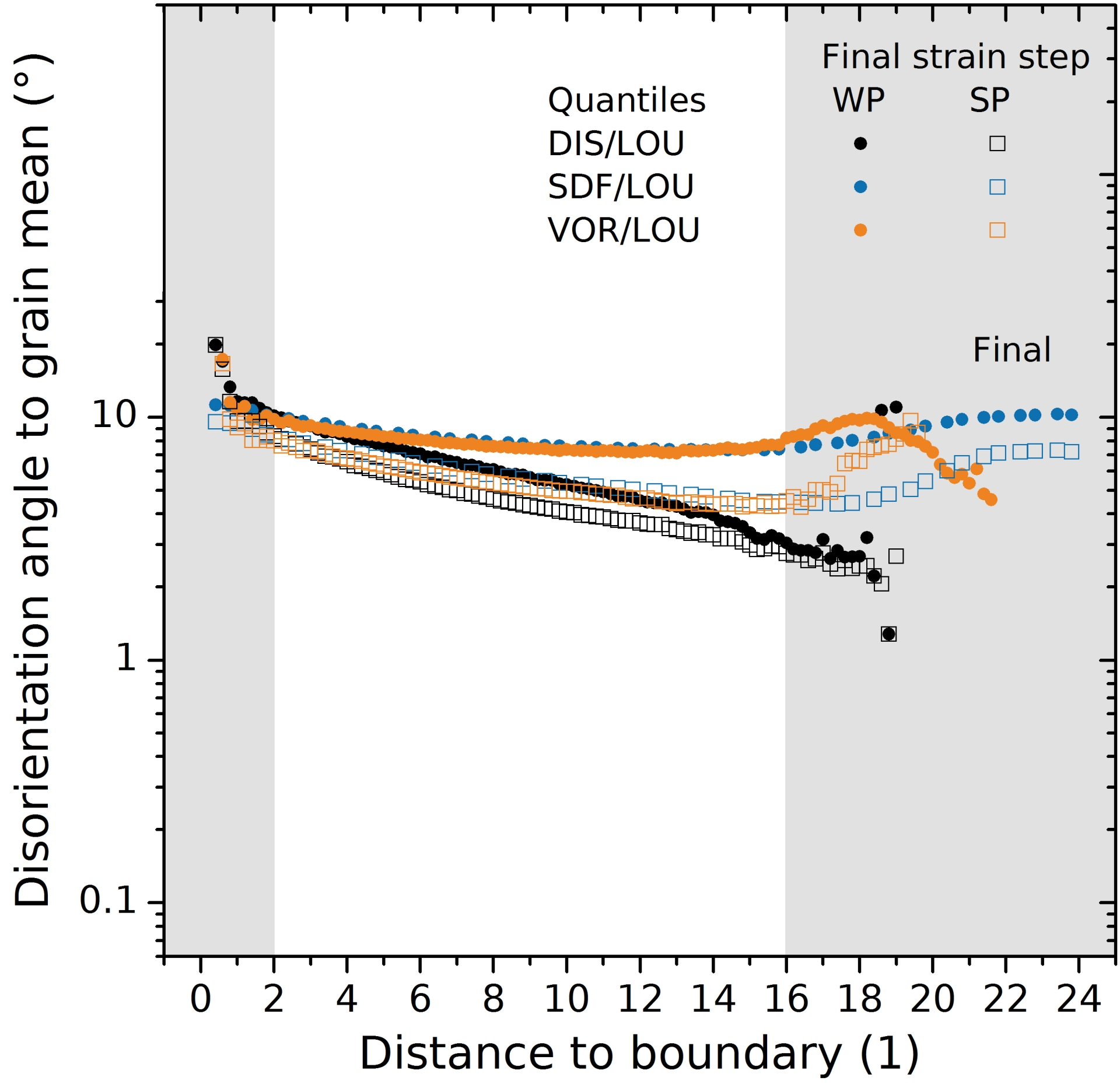}\label{DisoriDistroALLLOU}}
\caption{Key quantile values of the distribution of material point disorientation to the mean orientation of the reconstructed grain. Abbreviations discern like in Fig. \ref{StressDistrosAssessment} the distancing methods: disorientation (DIS), signed distance / voxelization (SDF), or tessellation (VOR) based, and how disorientations during grain reconstruction with the Louvain method were penalized (strongly SP or weakly WP). The distribution mean and the upper quantile (0.99) detail the spread of the distributions. In a) distributions are plotted at yield and final strain, while b) pulls focus on the final strain. As in Fig. \ref{StressDistrosAssessment} grayed-out regions detail were either numerical or finite counting effects are strongest.}
\label{DisoriVersusAllMethods}
\end{figure}

The results in Fig. \ref{DisoriVersusAllMethods} deliver quantitative evidence that spatial gradients of disorientation are detectable with all grain reconstruction and distancing methods: material points in the proximity of a grain boundary show on average larger disorientation to the mean orientation of the grain than do points in the grain interior. It can be excluded that these differences are the result of averaging orientations from multiple material points, as it is the case when measuring e.g. kernel average misorientations (KAM) at boundaries using the SEM/EBSD technique. In fact, when quantifying KAM values a kernel with multiple neighboring material points is evaluated. Upon probing with the kernel into neighboring grains this could contribute high disorientation values only if insufficiently strong criteria are set with respect to how much disorientation noise is allowed for a given kernel. 

Instead, in this work the disorientation values were computed independently to only one value for each material point - the respective grain mean orientation. In other words, no kernel was used. It can also be excluded that the observed gradients are random spatial correlations for the following reasoning: if a grain contains multiple material points in different orientations in the fundamental zone, it is expected that non-directionally correlated point-to-point disorientations are measured if one eventually compares the disorientation for two arbitrarily picked material points. If the orientation variation is strong and the grain small, i.e. the grain has few points support, eventually larger disorientations are measured on average. However, this scatter should also be correlated if there are orientation gradients with a strong component normal to the interface within the grain. Vice versa this scatter should remain practically uncorrelated if such gradients are absent or the grain contains practically uncorrelated small regions with spurious higher disorientation.

Given the strength of the gradient and mean disorientation value, one can conclude that the grains are in an incipient stage of fragmentation. They have not yet accumulated localized regions of significant point-to-point disorientation in the high-angle boundary regime (\SI{15}{\degree}), except for a minority population of material points in an at most $2$ pixel wide zone at the boundaries. Qualitatively, their higher disorientation is expected as in this zone also the absolute Cauchy stress values are also slightly higher (Figs. \ref{StressDistrosAssessment}).

There are two additional observations to make with practical significance for characterizing orientation gradients: first, only the methods with normal distancing capability (SDF, VOR) report similar gradient slope ($\approx \frac{\SI{4}{\degree}}{15px}$). Compared to disorientation based distancing, the slope is moderately lower. Second, disorientation gradient characterization with the graph clustering method shows a strong parameter sensitivity (Fig. \ref{DisoriDistroALLLOU}). 


\section{Discussion}
\subsection{Comparing the two grain reconstruction methods}
\begin{figure}[!ht]
\centering
	\subfloat[a][Number of reconstructed grains versus their size for a weak (WP) versus a strong (SP) penalization]{\includegraphics[width=0.45\textwidth]{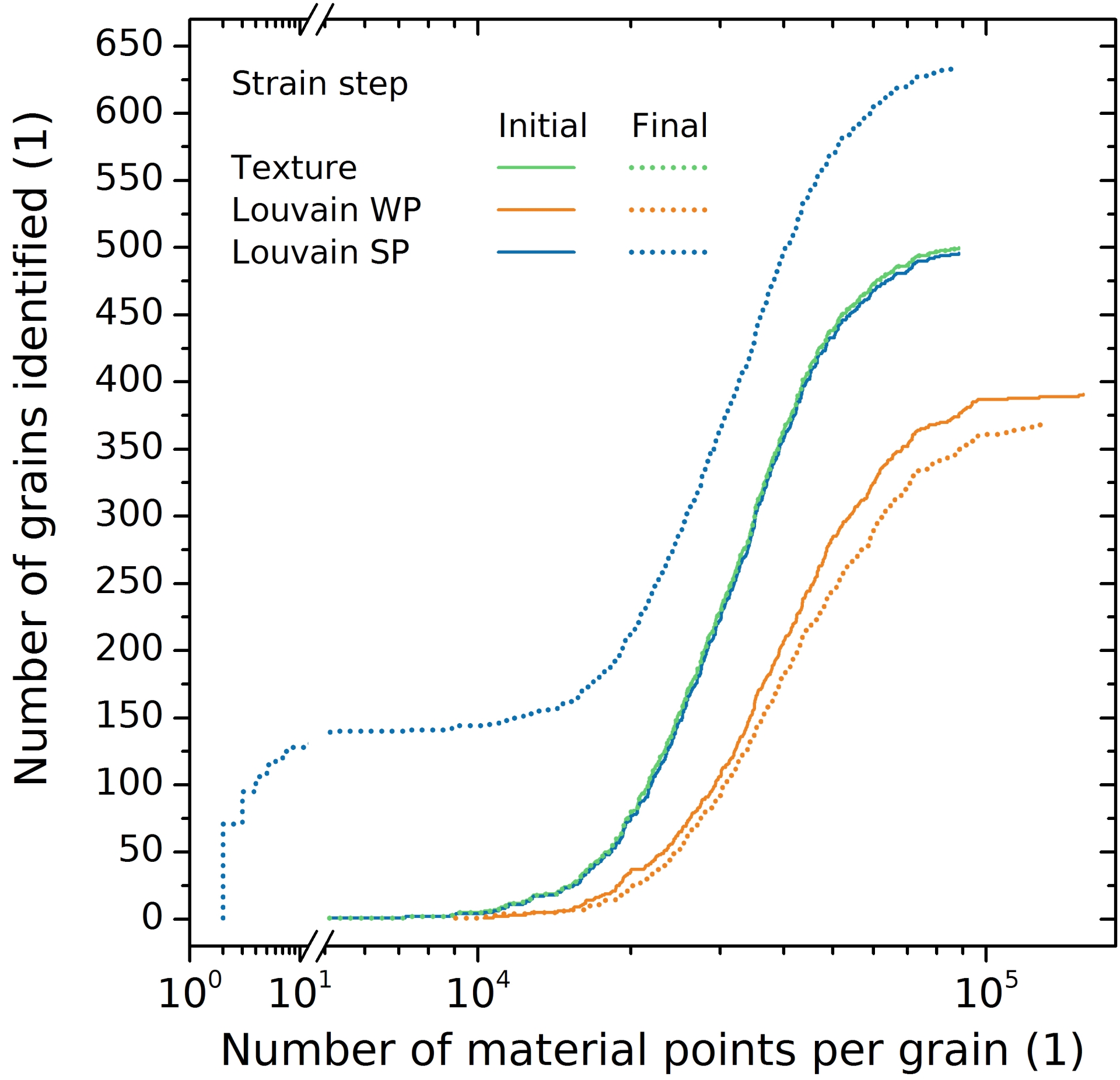}\label{GrainDistros}}
	\\
	\subfloat[b][A single grain resolved assessment of the graph clustering method \cite{Dancette2016} identifies that for a weak penalization $K_L = 75$ preferentially neighbors with low disorientation are merged, thereby reducing the total number of grains.]{\includegraphics[width=0.9\textwidth]{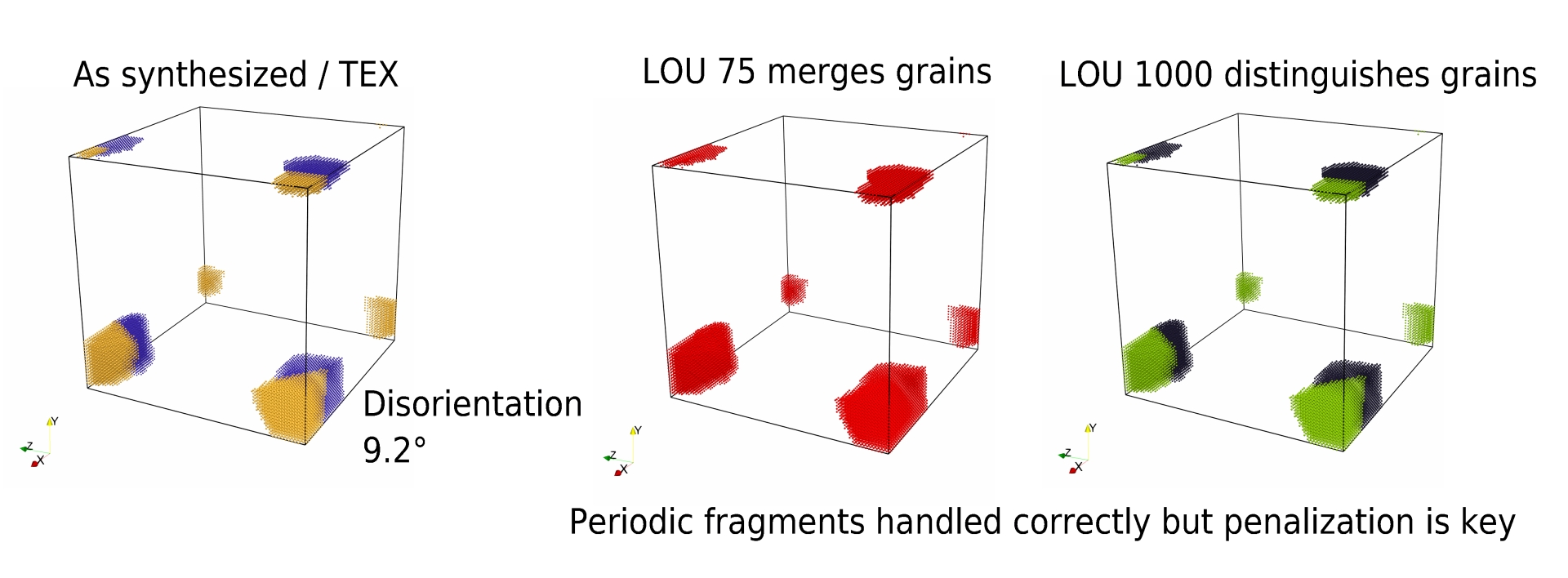}\label{ExemplarFusing}}
\caption{Texture ID versus graph clustering grain reconstruction: compared in terms of how many grains were reconstructed and how individual grains were (re-)identifiable for different penalization. Initial versus the final strain step results are compared.}
\label{LouvainAssessment}
\end{figure}

We identified that one practical challenge of using the graph clustering method for grain reconstructing is the strong parameter sensitivity. Therefore, it was worth studying the grain size distributions (Fig. \ref{GrainDistros}) and consistency with which individual grains were (re-)identifiable (Fig. \ref{ExemplarFusing}) as a function of the penalization parameter $K_L$ (algorithm \ref{AlgoLOU}). Figure \ref{LouvainAssessment} summarizes the key findings.

The larger the penalization parameter $K_L$ is set the stronger every accumulation of high disorientation between nodes of the same community gets penalized. In effect, more grains with lower in-grain orientation variation are found on average for a stronger penalization ($K_L = 1000$). 

Neither the values for a low nor for a high penalization can avoid systematic challenges when using the reconstructed grain ensemble for characterizing the distance correlations. In fact, if the penalization is weak, the algorithm reconstructs different grains and comes not even close to the initially synthesized number. Figure \ref{ExemplarFusing} proofs that this is a systematic consequence of the method's tendency to merge neighboring grains with low disorientation at low $K_L$ values. 

In effect, the mean orientation of many grain pairs is an average of at least two orientation ensembles with low intra ensemble but possible noticable inter ensemble disorientation. With respect to spatial disorientation gradients this has two consequences: first, such gradients are flatter and second they are shifted to larger disorientation on average. This explains why the gradients for weakly penalized graph clustering (Fig. \ref{DisoriDistroALLLOU}) are shallower.

One could avoid the possible bias introduced from the systematic grain merging by using strong penalization, such that eventually the initial grains are re-identified (Fig. \ref{GrainDistros}). Figure \ref{GrainDistros} shows the detrimental effect of such procedure when applying it to analyze the (highly) deformed configurations: oversegmentation with an associated qualitative change of the grain size distribution from unimodal to bimodal is observed. A detailed inspection of the respective material points identified that this spreading into eventually bimodality is a consequence of the fact that the strongest disoriented material points at the grain boundary get now segmented into individual small grains.



These findings have a methodological and a practical implication. The methodological implication is that graph clustering based grain reconstructions should always be backed up by rigorous quantitative parameter sensitivity studies. One strategy could be to report always unnormalized distributions of grain sizes as a function of the segmentation parameter. Otherwise, different physical conclusions might be drawn even though one uses the same method. One example, disorientation accumulates towards the boundaries versus it does not, is exemplified with Fig. \ref{DisoriDistroALLLOU} in this work. 

The practical implication reads as follows: grain fragmentation should be better characterized ideally via the evolution of the grain boundary network rather than to continue insisting logically on the initial grain as the decisive object. In fact, not the homogeneous regions within a deformed grain but the heterogeneities are the key microstructure locations and first descriptors to investigate where and how annealing microstructure evolution initiates.

\subsection{Implications of numerics-induced scatter for non-local and coupled crystal plasticity/interface migration spectral method models}
Evincing such numerics-induced scatter suggests to apply above quantification protocol in the future regularly. Especially, it should be applied in so-called non-local crystal plasticity models. These predict the local material point values via evaluating a kernel of neighboring integration / material points. Examples are continuum scale crystal plasticity solvers with incrementally and locally coupled dislocation flux sub-models, such as the one from \cite{Arsenlis2004,Reuber2014} or full-field crystal plasticity models with incremental coupling to phase-field solvers.

One should always start inspecting such solvers from an in-depth monitoring of the (state variable) values and the non-local flux terms for each material point as a function of strain. Methods, such as the ones developed in this paper, could be used to assess via such monitoring in how far numerical noise remains uncorrelated and low to avoid a systematic biasing of the results towards larger strains.

Another example in which rigorous quantitative monitoring of state variable values is useful are incrementally coupled grain boundary migration / crystal plasticity models. These typically evolve microstructures along complex transients. As such, any permeation of correlated numerical noise into the flux or interface migration terms should as best as possible be reduced to keep rigorously controlled conditions, physical accuracy and precision. 

Knowledge in the discrete dislocation dynamics community teaches us that uncontrolled stress spikes should better remain numerically controlled, for instance by supplementing the integration with some time trajectory before making choices \cite{VanderGiessen1994,Cleveringa1998}. For these future challenges the work delivers an approach based on which to build further tools for assessing the numerical and physical quality of such predictions.

\subsection{Comparing computational efficiency}
\begin{figure}[!ht]
\centering
	\includegraphics[width=1.0\textwidth]{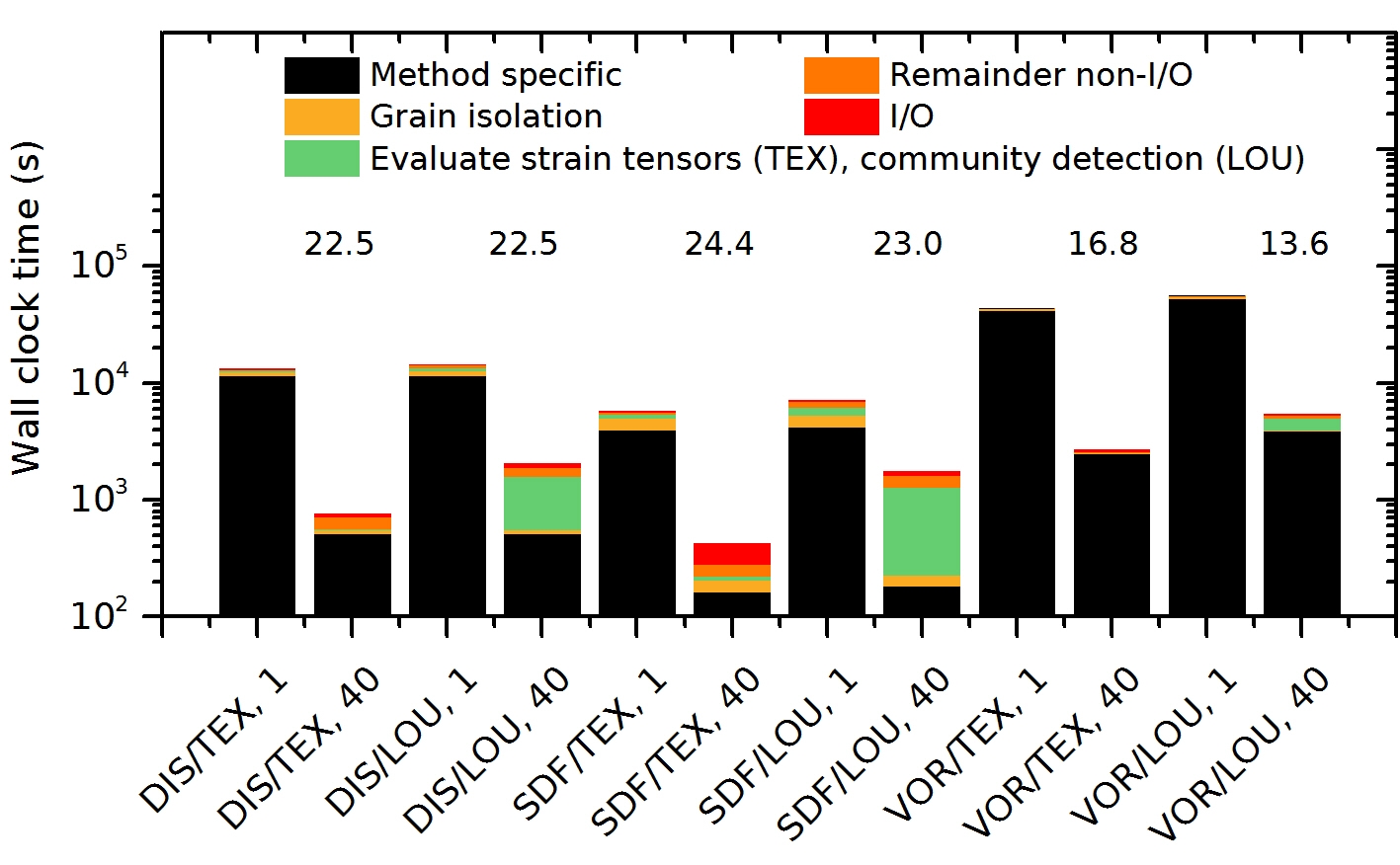}
\caption{Key results from benchmarking all three methods in combination with both grain reconstruction protocols using the same strain step (\ervat{0.21}) and the same cluster computer. The numbers above the right column of each column pair denote the respective speedup achieved for the most costly method specific processing step when executing in parallel. Abbreviations identify the disorientation based (DIS), the signed distance/voxelization based (SDF), and the tessellation based distancing (VOR). Grain reconstruction was based either on the initial assignment (i.e. texture ID based TEX) or graph clustering (LOU). Sequential execution is compared with threaded execution. Analyses using the LOU method have, different than those for TEX, a different sequential bottleneck (indicated e.g. for DIS/LOU, 40 by the green column portion): the community detection stage, which was in this work not parallelized.}
\label{AllMethodsHeadsUp}
\end{figure}

\paragraph{Strong scaling multithreading performance}
To document the strong scaling efficiency of the multithreading capabilities, Fig. \ref{AllMethodsHeadsUp} summarizes wall clock timings. Furthermore, the figure documents key computing time contributions for all combinations of above post-processing methods applied to the same data, using the same computer, and exclusive execution. The left column of each method column pair in Fig. \ref{AllMethodsHeadsUp} documents the sequential execution time. The results identify that signed distance / voxelization based distancing (SDF), in combination with either of the two grain reconstruction methods, is the least costly combination. Results for a single strain step and sequential execution are available after \SI{2}{\hour} approximately. Given the capability to identify normal distances is another argument to choose the SDF method over disorientation based distancing. 

Figure \ref{AllMethodsHeadsUp} documents that all methods concentrate at least \SIrange{87.5}{99.3}{\percent} of the total execution time in not more than three algorithmic code sections. For the disorientation based distancing these steps are the querying of neighboring points and computing point-to-point disorientation angles. For the signed distancing method it is the identification of periodic images, volume rediscretization, and the remapping of every voxel to a closest unique material point. For tessellation based distancing the evaluation of the projected normal distances for each material point demands contour facet querying.

In this work, all these most costly steps are parallelizable. Thus, specific care was taken to implement them in parallel. In addition, data placement was orchestrated such that \SI{}{\giga\byte}-sized data chunks were placed close in the memory of the executing core taking into account the memory hierarchy of the cluster computer.

In effect, this speeds up the execution of every method combination by \SIrange{13.6}{24.4}{} times the sequential baseline. With $40$ OpenMP threads required to achieve this, a strong scaling efficiency between \SIrange{61}{34}{\percent} is documented. The main obstacle to achieve higher efficiency in this study was  load imbalance. Such was strongest for the tessellation (VOR) method because grains were distributed round-robin, i.e. statically to the threads. When building tessellations, however, the grain size critically affects how many facets the contour hull contains; and thus how dissimilar the distancing costs are between different grains. In light of this splitting a population with only $500$ grains across $40$ threads will equip the threads with possibly too few grains to compensate for the large variety of grain sizes, i.e. cost and facet count differences in the contour hull processing stage.

A possible improvement for the future is to change the implementation and employ dynamic work scheduling. A possible solution could be to use the OpenMP task construct to delegate the processing of the grains to a queue which the threads then machine off dynamically. However, this eventually demands for a compromise: the current round-robin grain distributing procedure allows one to place them in a controlled memory location \cite{Hennessy2012}. A dynamic scheduling, in turn, will typically keep the cores more frequently busy at the cost of typically more memory traffic. 

Another potential improvement of this work, with immediate practical benefit, is to parallelize also the graph clustering algorithm for which CPU- and GPU-parallelized community detection solvers have been developed in the past \cite{Chavarria2014,Lu2015,Naim2017,Ghosh2018}. 

The results document additional practical improvements in which the present work advances the field: the first is improved speed, when compared to hitherto reported values for \dm post-processing \cite{Roters2019}. This was achieved with sequential optimization of the file accessing strategy surplus employing parallelized I/O to cut I/O costs by at least an order of magnitude. 

As an additional strategy to improve sequential performance, state of the art linear algebra libraries were used to compute point-wise tensor quantities. Such improvement is of practical use not only as \dm employs it for solving an RVE but also during post-processing to compute e.g. a flow curve. As the second improvement, this work details how to combine these strategies with multithreaded execution surplus trivial data parallel processing of strain steps on multiple nodes of a cluster computer.


Having the possibility to construct contour hulls to each grain in the deformed configuration is the third improvement. On the one hand because it allows to quantify a volume for each grain not only by counting the number of material points it contains but by accumulating the volume of its supporting Voronoi cells. The resulting polygon mesh allows for volumetric rendering of the grain as a polyhedron, as exemplified in Fig. \ref{VORVis}. These may be useful to allow further analyses using other microstructure characterization tools like DREAM3D \cite{Groeber2014} or QUBE \cite{Konijnenberg2013}. 

\begin{figure}[!ht]
\centering
	\includegraphics[width=1.0\textwidth]{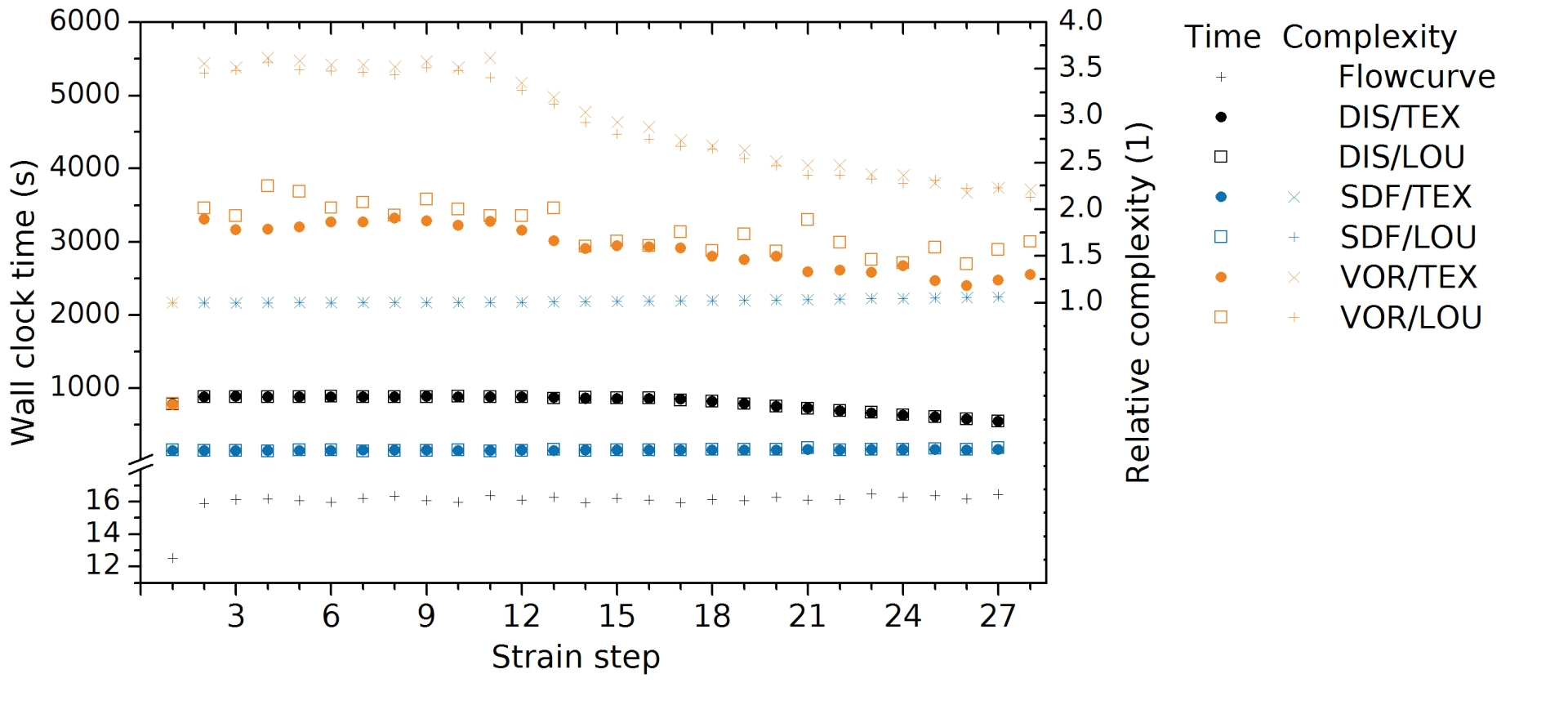}
\caption{Key results from benchmarking all three methods in combination with both grain reconstruction protocols on the entire strain step ensemble. Different computing costs are quantified via the wall clock time. The results identify that the amount of computational work is different across the strain step ensemble.
Taking the first strain step as the reference, relative complexities were established through monitoring e.g. how many voxel were processed for the ensemble of grain bounding boxes and how many Voronoi facet polygon inclusion and edge comparison tests were required to find the closest distances.}
\label{WeakScalingHeadsUp}
\end{figure}

\paragraph{Hybrid execution performance}
During hybrid execution every MPI process evaluates a different strain step and spawns an own group of threads. Data are read independently from a priori known sections of the \dm results file. In effect, this enabled to evaluate all $28$ strain steps at once using $1120$ cores. Figure \ref{WeakScalingHeadsUp} summarizes the key results for this hybrid execution mode. Specifically, two sets of common analysis cases were probed: a flow curve extraction versus analyses of aforementioned state variable values to boundary distance correlations, six of them in total.

There are two key findings of practical importance.
First, hybrid processing enables to cut wall clock time by an additional order of magnitude.
In effect, the evaluation of a flow curve, which took damaskpdt \SI{3.2}{\hour} sequentially, was solvable in \SI{76}{\second} when using parallelism. Similarly, distancing results for all strain steps were analyzed after \SI{1.5}{\hour} when using tessellation (VOR), the most costly method. Distances were even faster accessible, namely in \SI{8}{\minute}, using the signed distance method. 

The practical benefit of these findings is put in perspective by the following observation. While the SDF method was already finished for the entire ensemble, the classical \dm post-processing routines had not even completed the processing of the flow curve for the first strain step.

The second key finding is that parallel performance gains are limited. Respective arguments for the multithreaded analyses were already identified (Fig. \ref{AllMethodsHeadsUp}). Not parallelizing all post-processing computational steps is another apparent reason for limited scalability \cite{Amdahl1967}. 
The most significant obstacle to cut execution time further, though, are microstructure-induced differences. 
Specifically, differences which are caused by dissimilar interface geometry networks across the strain step ensemble. Consequently, the individual algorithms show different solving time resulting in workload differences across the strain step ensemble. 

Two observations are important to add here. First, these inter strain step work load differences would remain even if the multithreaded execution of each strain step gets better work load partitioned. Second, it is in fact the successive reduction of the relative numerical complexity for a strain step with increasing strain which limits primarily the hybrid performance of MPI parallelization. Provided there were more strain steps to process very likely even more processes could be employed. This is one of the key advantages to our true scientific computing performance solution for post-processing \dm simulations. 

One example is visible for signed distance based distancing. Given that the RVE shape change results in a moderate increase in the total bounding box volume, more voxel have to be probed for the higher than the lower strain steps. Another, by far stronger, example for microstructure induced work load imbalances is visible for tessellation based distancing (Fig. \ref{WeakScalingHeadsUp}).

When using the strongly penalized grain reconstruction, more grains are detected at higher strains. Therefore, also more boundary area is generated on average, as a consequence of which the material points lay closer to a boundary, providing for faster distancing operations. In addition also different contour hull shape and facet compositions are generated. This modifies the composition of the individual bounded volume hierarchies, and thus their individual response and pruning effectiveness during the querying stage. 

It is useful to compare the computational costs of post-processing to the actual simulation costs. The \dm deformation simulation occupied $36$ cores for \SI{354}{\hour} wall clock time, i.e. $12744$ core hours were spent in total. The post-processing of the entire strain step ensemble using the most costly distancing method kept $1120$ cores busy for \SI{4822}{\second} wall clock time, i.e. $1500$ core hours in total, or at most \SI{11.8}{\percent} of the simulation costs.

Lastly, the memory consumption of the simulation and post-processing should be reported. \dm allocated a total of \SI{230}{\giga\byte} virtual memory, i.e. \SI{14.3}{\kilo\byte} per material point. Post-processing demands less virtual memory per process when evaluating the flow curve only (\SI{21}{\giga\byte}). However, virtual memory consumption peaks at \SI{134}{\giga\byte} for the most costly strain step and using the VOR/LOU distancing method. One reason for the larger memory consumption are the storage demands for position values for a portion of the 26 periodic images to all material points. Another reason is that Voronoi polyhedra and their facets have to be stored for tessellation based grain reconstruction. 

Another contribution is due to the conservative strategy we used to implement the querying structures. This could be optimized in the future with potential for a memory footprint reduction of at least a factor two, thereby making possible also to post-process as large RVE simulation data on cluster computers with individually smaller node main memory than we used in this work.








\section{Conclusions}
A set of strong and weak scaling post-processing methods were developed to quantify the accumulation of state variable values at grain boundaries in 3D full-field spectral method crystal plasticity simulations. Exemplified for the \dm spectral solver and a phenomenological constitutive crystal plasticity model the results substantiate:
\begin{itemize}
    \item The methods successfully reconstruct the grain boundaries and operate on arbitrarily deformed and periodically confined RVE domains with respect to the deformed configuration. This allows for direct comparison to experiments and rigorous numerical assessment of the spectral method.
    \item Long range gradients of material point disorientation, with higher values to the boundary than in the grain interior, were quantified. They are a signature of incipient grain fragmentation.
    \item As an alternative method to capture the individual fragments, also a graph clustering grain reconstruction method was assessed. Compared to the classical method of assigning grain IDs based on the initial conditions, though, the quantification of the gradients is very parameter dependent. 
    \item Three methods to quantify the distance of a material point to the boundary were detailed. In effect, a combination of signed distance function representation and volume rediscretization delivered the lowest method affected quantification differences. Furthermore, such approach has the lowest numerical costs of all methods tested.
    \item All performance critical computational tasks were implemented distributed and shared memory parallel. Therewith, processing the entire set of $28$ strain steps from a $256^3$ RVE cube \dm simulation, worth \SI{162}{\giga\byte}, was possible in \SI{470}{\second} wall clock time using $28$ processes each spawning $40$ threads. Compared to exclusive multithreaded (\SI{11063}{\second}) or sequential execution (\SI{150000}{\second}), this hybrid approach allowed for $320$ times faster post-processing. The efficiency could be improved further if heterogeneous work load across the individual strain step data, which is method dependent, gets better balanced and remaining sequential code portions parallelized.
\end{itemize}

\section*{Data and code availability}
The source code of the tool and its supplementary \matlab scripts with which we post-processed all results are open source software. They are publicly accessible \cite{DAMASKPDT2019}. \debug{All \dm and post-processing parameter settings files, including the SLURM batch system submission scripts are available for download \cite{Kuehbach2019}\footnote{They will be made available upon acceptance of the article}}. The entire repository of compressed post-processed data occupies \SI{844}{\giga\byte}. It is available from the authors upon serious request.

\section*{Acknowledgements}
The authors gratefully acknowledge the funding received from the German Research Foundation through project RO 2342/8-1. The work was partially supported by BiGmax, the Max Planck Society's Research Network on Big-Data-Driven Materials-Science. MK acknowledges the discussions with Karo Sedighiani, Muhammad Imran, and Markus Bambach on parameterizing constitutive models, with Matthew Kasemer on implementing the stress conventions, and Martin Diehl for discussing about \dm in general and its post-processing scripts in particular.

\section*{Work distribution}
MK designed and implemented the tool surplus performed and post-processed the simulations. MK and FR wrote the manuscript and discussed the results on a regular basis.

\appendix
\section*{Appendices}
\addcontentsline{toc}{section}{Appendices}
\renewcommand{\thesubsection}{\Alph{subsection}}
\subsection{Detailed specification of the post-processing pipeline}
\paragraph{Algorithmic details}
\label{AlgoBasic}
In what follows, the individual steps of the post-processing pipeline are detailed. For all steps with a larger than linear time complexity $\mathcal{O}(N_xN_yN_z := N)$, additional details are commented. In every step, $N$ denotes the total number of unique material points supporting the RVE. Indices $n$ specify the strain step. Three point cloud sets are distinguished: the first contains the $N$ unique material points of the original domain \mNull $\in {\mathcal{R}}^3$. The second, \mNulleps, contains \mNull surplus all periodic images of \mNull which lay inside a cuboidal bounding box about a cuboidal tight bounding box to \mNull. The outer bounding box is fattened by $\epsilon = 0.1$ in length units of the initial RVE domain. The third point cloud set, \mOne contains $N_n^1$ points, i.e. all members of \mNull surplus all their 26 periodic copies. The members $i$ of the material point ensemble \mNull have positions $p_{i,n}^0$ which are defined by their initial locations in the RVE in the reference configuration \cite{Diehl2016,Roters2019} $x_{i,n}^0$ surplus a deformation-induced displacement which maps into the deformed configuration in the laboratory coordinate system $\Delta x_{i,n}^0$. Every material point has associated state variable values $s_{i,n}$. In addition, a reference ID and periodic image variant ID is stored to track which periodic image points are assigned to which unique material point of \mNull. This allows dereferencing of state variable values $s_{i,n}$ rather than duplicating them. IMKL specifies in which steps the Intel Math Kernel Library is used.

\begin{minipage}[]{0.95\textwidth}
\null 
 \begin{algorithm}[H]
    \caption{Post-processing pipeline}
    \begin{algorithmic}[1]
	\Procedure{{damaskpdt}}{}
		\State I/O parse parameter and initialize MPI
		\State I/O parse spectralOut file layout to identify content
		\State Partition $N_n$ strain increments round robin on MPI processes
		\For{$\epsilon_n \gets 1, \, N_n$}
		    \State Process parallelized MPI I/O spectralOut file reading
		    \State \Comment{State variable values $s_{i,n}$ per point $\bm{x, F_T, F_P, P, q}$}
		    \State Spatial distributing of material point cloud to OpenMP threads
		    \State Threaded stresses and strains $\forall i \in$ \mNull via IMKL
			\State RVE averaging $\overline{\bm{F}}, \overline{\bm{P}}, \overline{\sigma_{vM}}, \overline{\epsilon_{vM}}$
			\State Threaded displacements $\Delta \bm{x_{i,n}^0}$ via IMKL \Comment{$\mathcal{O}(Nlog(N))$}
			\State Compute periodic images \mNulleps
			\State Build a spatial index ${\mathcal{B}}_n^{0+\epsilon}$ that partitions \mNulleps
			\State{Threaded grain reconstruction}
    			\If {Initial assignment / texture ID based}
    			    \State RECONSTRUCT\_GRAINS\_TEXTUREID (see \ref{AlgoTEX})
    			\EndIf
    			\If {Community detection based}
    			    \State RECONSTRUCT\_GRAINS\_LOUVAIN (see \ref{AlgoLOU})
    			\EndIf
			\State{GRAIN\_GEOMETRY\_SIMPLIFICATION (see \ref{AlgoVOXEL})}
			\State{Threaded state variable/distance to boundary quantification}
    			\If {Disorientation thresholding based scalar distances}
    				\State ANALYZE\_DIS (see \ref{AlgoDIS})
    			\EndIf
    			\If {Signed distance / voxelization based normal distances}
    				\State ANALYZE\_SDF (see \ref{AlgoSDF})
    			\EndIf
    			\If {Tessellation based contour hull normal distances}
    				\State ANALYZE\_VOR (see \ref{AlgoVOR})	
    			\EndIf
			\State MPI parallel I/O binary results ${(d,\bm{\sigma, q})}_{i,n} \forall$ voxel or $i$
			\State MPI aggregate flow curve data ${(\epsilon_{vM}, \sigma_{vM})}_n$
		\EndFor
		\State I/O flow curve, internal profiling results, finalize MPI and exit
	\EndProcedure
	\end{algorithmic}
  \end{algorithm}
\end{minipage}

\subsection{Detailed specification of the grain reconstruction methods}
\paragraph{Texture ID based grain reconstruction}
\label{AlgoTEX}
${\mathcal{T}}_i$ specifies the initial texture ID assigned during microstructure synthesis \dm \cite{Diehl2016,Roters2019}. \\

\begin{minipage}[]{0.95\textwidth}
\null 
 \begin{algorithm}[H]
    \caption{Texture based grain reconstruction}
    \begin{algorithmic}[1]
	\Procedure{{RECONSTRUCT\_GRAINS\_TEXTUREID}}{}
		\For{$i \gets 1, \, N$}
			\State Assign point $i$ a grain ID ${\mathcal{G}}_{i,n} := {\mathcal{T}}_i$
		\EndFor
	\EndProcedure
	\end{algorithmic}
  \end{algorithm}
\end{minipage}

\paragraph{Graph clustering based grain reconstruction}
\label{AlgoLOU}
$N_{mv}$ specifies the number of executed node relabeling operations in the current iteration step, $Q$ the Newman-Girvan modularity \cite{Newman2004} using a critical quality value of $Q_c = \SI[mode=math]{0.01}{}$. $K_L$ specifies the penalization strength, i.e. whether a weak $K_L = 75$ or a strong $K_L = 1000$ penalization was used. $\Vert {[q_{i,j,n}]}_0 \Vert$ denotes the rotation angle argument of the disorientation quaternion. The search radius during graph edge construction was chosen as $R^{lv}_c = \frac{2}{256}$ distance units. \\
\begin{minipage}[]{0.95\textwidth}
\null 
 \begin{algorithm}[H]
    \caption{Louvain community based grain reconstruction}
    \begin{algorithmic}[1]
	\Procedure{{RECONSTRUCT\_GRAINS\_LOUVAIN}}{}
		\For{$i \gets 1, \, N$}
		    \State Find all neighboring points $p_{j,n} \in$ \mNulleps $\mid \{ \Vert \bm{x_{i,n}} - \bm{x_{j,n}} \Vert \leq R^{lv}_c \}$
		    \State Compute disorientation angle ${[q_{i,j,n}]}_0$
		    \State Compute edge weights $w_{i,j,n} = exp(K_L \cdot (\Vert {[q_{i,j,n}]}_0 \Vert - 1))$
			\State Add one graph edge with weight $w_{i,j,n}$ for every position pair $p_{i,n}$, $p_{j,n}$
		\EndFor
		\Do
			\State Iterative graph clustering according to \cite{Dancette2016,Blondel2008,Browet2011,Blondel2015}
  		\doWhile {$N_{mv} > 0$ and $Q > Q_c$}
  		\For{$i \gets 1, \, N$}
			\State Assign $i$ the final label of the top-level community as grain ID
		\EndFor
	\EndProcedure
	\end{algorithmic}
  \end{algorithm}
\end{minipage}

\paragraph{Grain geometry simplification}
\label{AlgoVOXEL}
$N^0_{k,n}$ and $N^1_{k,n}$, respectively are the number of material points supporting the $k$-th grain in \mNull and \mOne respectively. $A_{k,n}$ specifies a tight axis-aligned bounding box about the positions $N^1_{k,n}$. ${\mathcal{A}}_n$ is the tight cuboidal global bounding box which contains all $A_{k,n}$. In this work a cell size of half the normalized initial material point point-to-point distance was used, i.e. $d_{cell} := \frac{1}{2\cdot256}$.
\begin{minipage}[]{0.95\textwidth}
\null
 \begin{algorithm}[H]
    \caption{Threaded extraction of a single periodic image per grain}
    \begin{algorithmic}[1]
	\Procedure{{GRAIN\_GEOMETRY\_SIMPLIFICATION}}{}
		\State Re-organize members of \mNull into disjoint sub-sets \kNull per grain $k$
		\Comment{In what follows, multithreaded processing $\forall k$ grains}
		\State Threaded memory initialization
		\State Threaded periodic images \kOne and bounding boxes ${\mathcal{A}}_k$
		\State Threaded building of sparse spatial index ${\mathcal{B}}^1_{k,n}$ via ${\mathcal{B}}_n^1$
		\State Threaded merging of possible grain fragments via DBScan \cite{Ester1996} 
		\Comment{$\mathcal{O}(N_klog(N_k))$}
		\State Threaded picking of one representative per grain	${\hat{\mathcal{A}}_{k,n}}$
		\State Identify global bounding box ${\mathcal{A}}_n$ enclosing all ${\hat{\mathcal{A}}_{k,n}}$
		\State Define a global voxelization ${\mathcal{L}}_n$ of volume ${\mathcal{A}}_n$ with cell size $d_{cell}$
		\State Fuse members of ${\mathcal{B}}^1_{k,n}$ into one global spatial index ${\mathcal{B}}_n^1$
	\EndProcedure
	\end{algorithmic}
  \end{algorithm}
\end{minipage}

\subsection{Detailed specification of the distancing methods}
\paragraph{Disorientation based distancing}
\label{AlgoDIS}
A disorientation angle of $\Theta_c = \SI{15}{\degree}$ was used for thresholding. \\
\begin{minipage}[]{0.95\textwidth}
\null
 \begin{algorithm}[H]
    \caption{Disorientation thresholding based scalar distances}
    \begin{algorithmic}[1]
	\Procedure{{ANALYZE\_DIS}}{}
		\For{$i \gets 1, \, N$}
		    \Comment{Multithreaded, dynamic scheduling}
			\State Find all $N_j$ neighbors $p^{0+\epsilon}_j \in$ \mNulleps $\mid \{ \bm{x_{i,n}} - \bm{x_{j,n}} \| \leq R^{lv}_c \}$
			\State Sort by distances $d_{i,j,n} := \Vert \bm{x_{i,n}} - \bm{x_{j,n}} \Vert$								\Comment{$\mathcal{O}(N_jlog(N_j))$}
			\For{$j \gets 1, \, N_j$}
				\State Compute disorientation angle $\Theta_{i,j,n}$	
				\If {$\Theta_{i,j,n} \geq \Theta_c$}
					\State Report $(d_{i,j,n}, \bm{s_{i,n}})$
					\State break		
				\EndIf
			\EndFor																\Comment{ignore $i$ if no pair found}			
		\EndFor
	\EndProcedure
	\end{algorithmic}
  \end{algorithm}
\end{minipage}

\paragraph{Signed distance / voxelization based normal distances}
\label{AlgoSDF}
Abbreviations denote a signed distance function (SDF) and the fast sweeping method (FSM) \cite{Zhao2004,Miessen2017}. 
\begin{minipage}[]{0.95\textwidth}
\null
 \begin{algorithm}[H]
    \caption{Signed distance / voxelization based normal distances}
    \begin{algorithmic}[2]
	\Procedure{{ANALYZE\_SDF}}{}
		\State Threaded voxelizing of $k$ according to ${\mathcal{L}}_n$, $d_{cell}$ using ${\mathcal{B}}_n^1$
		\State Specifically, $\forall$ voxel $\in {\mathcal{B}}_{k,n}^1$ identify closest member of \mNull
		\State Threaded initialization of SDF $\Phi_{k,n}(x)_0$
		\State Threaded spreading of SDF via FSM $\Phi_{k,n}(x)_1$ as in \cite{Miessen2017}
		\State Threaded report $(\Phi_{k,n}(x), \bm{s_{j,n}})$ $\forall$ voxel with $\Phi_{k,n}(x) \geq 0$
	\EndProcedure
	\end{algorithmic}
  \end{algorithm}
\end{minipage}

\subsection{Tessellation based contour hull normal distances}
\label{AlgoVOR}
${\hat{\mathcal{A}}}_{k,n}$ specifies an axis-aligned bounding box about the identified single periodic image of grain $k$. The box is fattened by a guard zone of thickness $\frac{3}{256}$ of the normalized initial material point-to-point distance. The fattening assures that the volume about each material point belonging to the grain $k$ can be tessellated and no Voronoi cell gets arbitrarily cut off by the bounding box walls. The query structure of the \voroxx library was configured to include five points per spatial bin on average. For each grain $k$ interior points are distinguished from exterior points with respect to their logical identity when building the exterior contour hull of the grain. Specifically, interior points $ip$ originate from a material point in \mNull with grain id $k$. BVH is short for bounded volume hierarchy. $N^k_f$ denotes the number of exterior facets identified for each grain $k$. $N^k_{ip}$ is the number of interior points per grain $k$. Exterior facets to a Voronoi cell of an interior point are all facets to the first order neighboring cells of an exterior point.
\begin{minipage}[]{0.95\textwidth}
\null
 \begin{algorithm}[H]
    \caption{Tessellation based contour hull normal distances}
    \begin{algorithmic}[1]
	\Procedure{{ANALYZE\_VOR}}{}
	    \Comment{$\forall k$}
		\State Threaded pulling of all points ${\hat{\mathcal{A}}}_{k,n}$ from ${\mathcal{B}}_n^1$
		\State Threaded mapping periodic image point to unique material point 
		\State Threaded \voroxx build spatial index to accelerate tessellator
		\State Threaded \voroxx tessellate all Voronoi cells for points
		\State Threaded \voroxx exterior facets, consistent outer unit normal
		\Procedure{IDENTIFY\_NORMAL\_DISTANCES}{}
		    \State Threaded construction of a BVH with all facet polygon
		    \For{$ip \gets 1, \, N^k_{ip}$}
		        \State Pool all $nf$ facets $f$ of the contour as initial candidates
		        \While{$f < nf$}
		            \State Normal projection interior point $x_{in}$ on facet plane $x^f_{proj}$
		            \If {$x_{proj}$ covered by facet polygon $=$ false}
		                \For{$e \gets 1, \, N^f_{e}$} \Comment{Circulate facet edges $e$ of $f$}
		                    \State Normal projection $x_{in}$ on $e$
		                    \If {$x^e_{proj}$ on the edge $=$ true}
		                        \State Eventually update current shortest distance
		                        \State Eventually re-query facet candidate pool
		                    \EndIf
		                \EndFor
		                \For{$v \gets 1, \, N^f_{v}$} \Comment{Circulate facet vertices $v$ of $f$}
		                    \State Compute distance $d_{in,v} = \| x_{in} - x_v \|$
		                    \State Eventually update current shortest distance
		                    \State Eventually re-query facet candidate pool
		               \EndFor
		            \Else
		                \State Eventually update current shortest distance
		                \State Eventually re-query facet candidate pool
		            \EndIf
                \EndWhile
		    \EndFor
		\EndProcedure
		\State Threaded report value pairs $(d_{ip,n}, \bm{s_{ip,n}})$ $\forall ip$
	\EndProcedure
	\end{algorithmic}
  \end{algorithm}
\end{minipage}

\section*{References}
\bibliography{drxnuc_chargradients_05}

\end{document}